 \documentclass[twocolumn,twocolappendix]{aastex631}

\usepackage{multirow}
\usepackage{booktabs}
\usepackage{newtxtext,newtxmath,ulem}
\usepackage[T1]{fontenc}
\usepackage{amsmath}
\usepackage{xcolor}
\usepackage{mathtools}

\newcommand{\Msun}{\ensuremath{\mathrm{M}_\odot}}

\def\be{\begin{equation}}
\def\ee{\end{equation}}
\def\ba{\begin{eqnarray}}
\def\ea{\end{eqnarray}}

\def\ltsima{$\; \buildrel < \over \sim \;$}
\def\simlt{\lower.5ex\hbox{\ltsima}}
\def\gtsima{$\; \buildrel > \over \sim \;$}
\def\simgt{\lower.5ex\hbox{\gtsima}}

\definecolor{falured}{rgb}{0.5, 0.09, 0.09}

\shorttitle{Super-virial gas in absorption}
\shortauthors{Bisht et al.}

\begin{document}

\title{Can supernova from runaway stars mimic the signs of absorbing `super-virial' gas? }

\correspondingauthor{Mukesh Singh Bisht}
\email{msbisht@rrimail.rri.res.in}

\author[0000-0002-1497-4645]{Mukesh Singh Bisht}
\affiliation{Raman Research Institute
Bengaluru - 560080, India}

\author[0000-0002-6389-2697]{Projjwal Banerjee}
\affiliation{Department of Physics,  Indian Institute of Technology Palakkad, Kerala, India}

\author[0000-0003-1922-9406]{Biman B. Nath}
\affiliation{Raman Research Institute
Bengaluru - 560080, India}

\author[0000-0002-3463-7339]{Yuri Shchekinov}
\affiliation{Raman Research Institute
Bengaluru - 560080, India}



\begin{abstract}
The recent detection of large column density absorption lines from highly ionized gas in a few directions through the circumgalactic medium (CGM) of the Milky Way (MW) has been puzzling. The inferred temperature from these absorption lines far exceeds the virial temperature of the MW, and the column densities are also too large to be easily explained. In this paper, we propose a novel idea to explain these observations and claim that they may not have originated from the CGM, but from a totally different type of source, namely, stellar ejecta from supernovae (SNe) above the Galactic disk that happen to lie in the line of sight to the background quasars.
About $\sim 20\%$ of massive OB stars (progenitors of core-collapse supernovae) are known to be runaway stars that have high ejection velocities near the Galactic plane and can end up exploding as SNe above the Galactic disk.
We show that the associated reverse shock in the supernova remnant in the early non-radiative phase can heat the ejecta to temperatures of $\gtrsim 10^7\,{\rm K}$ and can naturally explain the observed high column density of ions in the observed `super-virial' phase along with $\alpha$-enriched super-solar abundance that is typical of core-collapse supernovae. However, SNe from runaway stars has a covering fraction of $\lesssim 0.7 \%$ and thus can only explain the observations along limited sightlines.

\end{abstract}



\section{introduction}
Circumgalactic medium (CGM) is the gaseous component surrounding the stellar disk and the interstellar medium (ISM) of the Galaxy (\citet{Tumlinson2017}, \citet{Faucher2023}). The gas in the CGM is multiphase, as suggested by observations and numerical simulations, and the gas temperature varies from $\sim 10^4$ K (cold phase), $\sim 10^5$ K (warm phase) to $\sim 10^6$ K (virial phase or hot phase). However, a few recent observations have reported the existence of an additional phase called the  `super-virial' phase ($\sim 10^7$ K) in the CGM of MW. 
 This has been detected by using absorption lines of Hydrogen and Helium-like ions of Si, Ne, and Mg in the X-ray spectra of bright quasars along various sight lines. \citet{Das2019a} used  {\it XMM-Newton} observations towards the Blazar IES 1553+113 ($l=21.91^{\circ}$, $b=43.96^{\circ}$) to discover this phase through NeIX and NeX ions in a rather high S/N spectra. They also detected other ions (see their Table 1) and found the super-virial phase to be $\alpha$-enhanced, with super-solar values of the ratios of ions (see their Table 2) indicating enrichment by core-collapse supernovae (CCSNe). 

 \begin{table*}
  \centering
 \setlength{\tabcolsep}{10pt}
    \renewcommand{\arraystretch}{1.5}
   \begin{tabular}{|c|c|c|c|c|}

     \hline
  
    \multirow{2}{4em}{Reference} & $N_{\rm OVIII}$ & $ N_{\rm NeX}$ & $N_{\rm SiXIV}$ \\\cline{2-4}

     & \multicolumn{3}{c|}{($\times 10^{15} \ \rm cm^{-2}$)} \\
     \hline
    \citet{Das2019a} (IES1553+113) & $3.4\pm 1.4$ & $15.8\pm 4.6$ &  -- \\
    \hline
    \citet{Das2021} (Mrk421) & $2.24\pm0.44$ & $2.12^{+0.70}_{-0.66}$ & $7.48\pm 2.08$ \\
    \hline
    \citet{McClain2023} (NGC3783) & $9.66^{+3.90}_{-3.93}$ & $15.28\pm 3.37$ &  $<35.26$ \\
    \hline
    
     \end{tabular}
     \caption{This table shows the observed column density of OVIII, NeX, and SiXIV from \citet{Das2019a}, \citet{Das2021}, and \citet{McClain2023} towards IES1553+113, Mrk421, and NGC3783 respectively.}
     \label{tab:data}
\end{table*} 

Further observation by \citet{Das2021} towards Mrk 421 ($l=179.83^{\circ}$, $b=65.03^{\circ}$) also detected the super-virial phase using Si and Mg ions. They detected a variety of other ions in the virial and lower temperature phases as well (see their Table 1). The super-virial phase was again found to be $\alpha$-enhanced, with super-solar values of the abundance ratio of ions (see their Table 2). In addition, they inferred the dominance of non-thermal broadening in this phase.  More recently, \citet{McClain2023} used {\it Chandra} observations towards the quasar NGC 3783 ($l=287.46^{\circ}$,  $b=22.95^{\circ}$) to study the gas in absorption. They detected the super-virial phase with Ne ions (see their Table 1). Stacking {\it Chandra} observations along 46 lines of sight \citet{Lara-DI2023} has also detected SiXIV and SXVI absorption lines with an average column density $8.7\times 10^{15}$ and $1.5\times 10^{16}\,{\rm cm^{-2}}$, respectively, with super-virial temperatures \citep{Lara-DI2024}.

Table \ref{tab:data} summarizes the absorption data for the SV absorbing phase from these observations for various ions. Observers had also modeled the absorption systems as consisting of gas at multiple temperatures, then used collisional ionization equilibrium (CIE) to deduce the temperature and ionic column densities of the multiple phases using the \textsc{phase} model (\citet{Krongold2003}). These quantities are modeled and derived inferences depend on various assumptions. In the following, we work with the column densities mentioned in Table \ref{tab:data} which are directly observed, and compare our model predictions with them. This is particularly important because in our proposed scenario that will be discussed later, CIE is not applicable and the super-viral nature of gas is implied due to high observed column densities of H-like ions such as OVIII, NeX, and SiXIV.
We will refer to gas at $T\gg 10^6$ K as the super-virial phase below. Note that He-like ions from these elements may have contributions from lower temperatures as well, which is why we focus on the H-like ions in this paper.

The signature of a $\sim 10^7\,{\rm K}$ gas phase has also been seen in X-ray emission over the entire sky and has been modeled as an additional component along with the virial phase (\citet{Das2019b}, \citet{Gupta2021}, \citet{Bluem2022}, \citet{Bhattacharyya2023}, \citet{Ponti2023} and \citet{Sugiyama2023}). The typical emission measure (EM) of this phase is of order $\approx 10^{-3}$ cm$^{-6}$ pc. The corresponding (equivalent Hydrogen) column density of this gas can be estimated as $N\approx (L \times EM)^{1/2}$, where $L$ is the size of the region occupied by hot gas. If this hot gas that is seen in emission {\it also} causes the absorption signatures, then the typical length scale would be $L \approx N^2/EM$. Using the reported column densities (of order $10^{21}$ cm$^{-2}$ (e.g., \citet{Das2021}) in absorption studies, the length scale turns out to be exceedingly large ($\gg$ Mpc). This implies that the $\sim 10^7\,{\rm K}$ gas that is seen in emission {\it cannot} explain the absorption observations. It is therefore reasonable to assume that the hot gas seen in emission and absorption have different origins. In this paper, we focus on an explanation of $T\gg 10^6$ K gas seen in absorption studies, and refer to it as `SV absorbing' phase.

We propose a novel explanation for the absorption line from the SV absorbing phase, and argue that it could be due to a supernova remnant (SNR) above the Galactic disk. Several observations have detected stars in the `extra-planar' region of the MW that is occupied by CGM. These stars are ejected from the Galactic disk through various processes. Stars can get a kick velocity if one of the stars in a binary system explodes as a supernova \citep{blaauw1961} or through dynamical interaction in large open clusters \citep{poveda1967}. These processes can launch a star into the CGM with speeds of $\lesssim 400\,{\rm km\, s^{-1}}$ which is less than the escape velocity. They are called runaway stars \citep{Silva2011}. The massive stars among these runaway stars can explode as CCSNe above the Galactic disk, in the extra-planar region. 

We demonstrate that the column density along observed lines of sight can be produced by extra-planar supernovae  if the lines of sight pass through them.  The reverse shock can heat up gas (shocked ejecta) to the super-virial temperatures ($\sim 10^7$ K). This reverse-shocked ejecta can explain the observed chemical properties of the super-virial phase, since it is $\alpha$-enriched and has a super-solar abundance ratio of ions \citep{Nomoto2006,woosley2007}. Our proposal, therefore, unburdens one from trying to explain the absorption lines in the context of MW-wide phenomena, the difficulty of which is illustrated by the estimates mentioned above. While the proposal is admittedly offbeat, our guiding principle is the old adage that after exhausting other possibilities, even an unusual idea, if deemed plausible, is worth pursuing. Our goal in this paper is to show the plausibility of the proposed scenario, along with offering a testable prediction.

Our paper is organized as follows. Section~\ref{sec:methods} describes the details of modeling of massive stars and core-collapse supernova and the associated evolution of the non-radiative supernova remnant, followed by results in Section~\ref{sec:results}. We discuss our results and its implications in Section~\ref{sec:discussion} and end with summary in Section \ref{sec:summary}.

\section{Simulation of Core-Collapse Supernova and Evolution of the Remnant}\label{sec:methods}
We first give a brief description of the simulation of the evolution of massive stars and the associated nucleosynthesis leading up to and including the CCSNe.
\subsection{Modelling Massive Star Evolution and Nucleosynthesis in CCSNe}

We use the 1D hydrodynamic code \textsc{kepler} \citep{weaver1978presupernova,rauscher2003hydrostatic} to simulate the detailed evolution of non-rotating massive stars with an initial solar composition adopted from \citet{Asplund2009}. We adopt the default procedures as the ones described in detail in \citet{woosley2007} with the detailed stellar physics described in \citet{woosley2002}. Briefly, we follow the evolution starting from the star's birth just before zero-age main sequence to the eventual end point of its life where the star explodes via CCSN. At all stages of evolution, the detailed nucleosynthesis is followed by a large adaptive co-processing network. Convective mixing is modelled using the mixing length theory with the Ledoux criteria for stability along with semiconvection and overshoot for convective boundaries \citep{woosley1988}. 
Mass loss via wind, which has a large impact on massive star evolution, is included with mass loss rates adapted from \citet{nieuwenhuijzen1990parametrization} for the main sequence and red giant phase. 
The explosion due to core-collapse supernova is modelled by driving a piston from the base of the mass shell where the entropy per baryon exceeds $4\,k_{\rm B}$ which roughly coincides with the base of the Oxygen shell \citep{woosley2007,heger2010nucleosynthesis}. The acceleration of the piston is adjusted so that the final kinetic energy matches the desired explosion energy. We consider models of mass ranging from $11$--$30\,\Msun$ with a fiducial explosion energy of $1.2\times 10^{51}\,{\rm erg}$. The amount of mass ejected during CCSN, $M_{\rm ej}$, depends on the mass left over following the mass lost in the wind. Since mass loss is stronger in more massive progenitors, lower mass models have very little mass loss whereas more massive models lose a substantial part of their mass. 
For example, the $11\,\Msun$ models loses only $\sim 0.35\,\Msun$ whereas a $25\,\Msun$ model loses $\sim 10\,\Msun$. The final values of $M_{\rm ej}$ varies between $\sim 9$--$13\,\Msun$. On the other hand, the mass of the core $M_{\rm c}$, defined as the part excluding the envelope where almost all the metals synthesized in the star are contained, also increases with the progenitor which varies between $1.25$--$8\,\Msun$.

\subsection{Modelling the Supernova Remnant}
We simulate the evolution of nonradiative SNR starting from the early ejecta-dominated (ED) phase to the Sedov-Taylor (ST) phase by using a simple 1D hydrodynamics code following the exact prescription of \citet{Truelove1999}. The code solves the Euler equations of hydrodynamics in Lagrangian form assuming spherical symmetry using a finite difference method for an equation of state of an ideal monoatomic gas ($\gamma=5/3$). The ambient medium is assumed to be cold (zero thermal pressure) and of uniform density $\rho_0$. The ejecta, following free expansion for a few years after the explosion, is also assumed be cold which is a very good approximation as indicated by \textsc{kepler} models. Instead of using an exact density profile from a freely expanding ejecta from the CCCN simulation discussed above, following \citet{Truelove1999}, we consider density profiles that do not introduce any length scales and allow us to write the Euler equations in terms of characteristic length, mass, and time scales such that the solution is dimensionless in characteristic units. The solution of the Euler equations in such cases is referred to as \emph{unified} solutions by \citep{Truelove1999}. The characteristic units depend on the total ejecta mass $M_{\rm ej}$, ambient density $\rho_0$, and the explosion energy $E$. Thus, for a given ejecta density profile, the solution for different values of $M_{\rm ej}$, $\rho_0$, and $E$ can be found from a single unique solution in terms of the characteristic units which are given by
\citep{Truelove1999};

\begin{align}
R_{\rm ch}&\equiv M_{\rm ej}^{1/3}\rho_0^{-1/3} \nonumber\\
&=14.73\left(\frac{M_{\rm ej} }{10\,\Msun} \right)^{1/3}  \left(\frac{{0.1\, \rm cm^{-3}}}{n_0}\right)^{1/3}~{\rm pc},\\
t_{\rm ch}&\equiv E^{-1/2} M_{\rm ej}^{5/6}\rho_0^{-1/3}\nonumber \\
&=6.43\left(\frac{{10^{51}\,{\rm erg}}}{E}\right)^{1/2}\left(\frac{M_{\rm ej} }{10\,\Msun} \right)^{5/6} \left(\frac{{0.1\, \rm cm^{-3}}}{n_0}\right)^{1/3}~{\rm kyr},\\
M_{\rm ch}&\equiv M_{\rm ej}\nonumber \\
&= 1.99\times10^{34} \left(\frac{M_{\rm ej} }{10\,\Msun} \right)~{\rm g},
\end{align}
where $n_0$ is the ambient atomic number density. We adopt a mean atomic number $\bar A=1.276$ for the ambient medium corresponding to a gas of solar composition form \citet{Asplund2009} such that $n_0=\rho_0/(\bar A m_{\rm H})$, where $m_{\rm H}$ is the mass of hydrogen atom. Based on the above characteristic quantities, we can also define the characteristic velocity $v_{\rm ch}$, density $\rho_{\rm ch}$, and pressure $P_{\rm ch}$ as
\begin{align}
    v_{\rm ch}&\equiv\frac{R_{\rm ch}}{t_{\rm ch}}=E^{1/2}M_{\rm ej}^{-1/2}\nonumber\\
              &=2.24\times 10^8 \left(\frac{E}{10^{51}\,{\rm erg}}\right)^{1/2}\left( \frac{10\,\Msun}{M_{\rm ej} }\right)^{1/2}~{\rm cm\, s^{-1}}\\
    \rho_{\rm ch} &\equiv \frac{M_{\rm ch}}{R_{\rm ch}^3}=\rho_0 \nonumber\\
                &= 2.12\times10^{-25} \left(\frac{n_0}{{0.1\,\rm cm^{-3}}}\right) ~{\rm g\,cm^{-3}}\\
    P_{\rm ch} &\equiv \frac{M_{\rm ch}}{R_{\rm ch}t_{\rm ch}^2}=EM_{\rm ej}^{-1}\rho_0 \nonumber \\
                &= 1.07\times10^{-8} \left(\frac{E}{10^{51}\,{\rm erg}}\right)\left( \frac{10\,\Msun}{M_{\rm ej} }\right) \left(\frac{n_0}{{0.1\,\rm cm^{-3}}}\right)\,{\rm dyne\,cm^{-2}}
\end{align}

\subsubsection{Initial velocity profiles}
The initial velocity profile is adopted assuming a freely expanding ejecta given by \citep{Truelove1999}
\begin{equation}
  v(r)=\begin{dcases}
    \frac{r}{t_{\rm ini}}, & \text{$r\leq R_{\rm ej}$}\\
    0, & \text{$r>R_{\rm ej}$}.
  \end{dcases}
\end{equation}
where $R_{\rm ej}$ is the radius of the expanding ejecta at the initial starting time $t_{\rm ini}$. This agrees very well with the simulated velocity profile from CCSN models from \textsc{kepler}.

\subsubsection{Density profile without a core}
For the density profile, we first consider an idealized ejecta with a constant density $\rho$ without any core, corresponding to the $n=0$ power-law density profile discussed in \citet{Truelove1999} given by 
\begin{equation}
  \rho(r)=\begin{dcases}
    \frac{3}{4\pi}\frac{M_{\rm ej}}{R_{\rm ej}^3}, & \text{$r\leq R_{\rm ej}$}\\
    \rho_0, & \text{$r>R_{\rm ej}$}.
  \end{dcases}
  \label{eq:rho_nocore}
\end{equation}
The value of $R_{\rm ej}$ for a given $t_{\rm ini}$ depends only on $M_{\rm ej}$ and $E$. In order to find this dependence, it is useful to define $v_{\rm ej}$ which is the maximum velocity of the ejecta corresponding to the outermost surface of the ejecta i.e., $v(r=R_{\rm ej})=R_{\rm ej}/t_{\rm ini}$. Using the fact that $E=0.5\int_0^{R_{\rm ej}}4\pi \rho r^2v^2dr$ we find 
\begin{align}
v_{\rm ej}&=\left(\frac{10}{3}\right)^{1/2}E^{1/2}M_{\rm ej}^{-1/2} \nonumber\\
        &=4.09\times 10^8 \left(\frac{E}{10^{51}~{\rm erg}}\right)^{1/2}\left( \frac{10\,\Msun}{M_{\rm ej}}\right)^{1/2}~{\rm cm\, s^{-1}}
\label{eq:vej_nocore}
\end{align}
Thus, for a given $t_{\rm ini}$, the velocity and density profile are uniquely determined by $M_{\rm ej}$ and $E$. Similar to \citet{Truelove1999}, we use the $*$ symbol to denote quantities in characteristic units such that $t^*=t/t_{\rm ch}$, $\rho^* = \rho/\rho_{\rm ch}$, etc. We adopt $t_{\rm ini}^*=0.01$ as the default value. In characteristic units, however, the velocity and density profiles are independent of $M_{\rm ej}$, $E$, and $n_0$ and only depend on $t_{\rm ini}^{*}$.  For the adopted density profile, we have a constant $v_{\rm ej}^{*}=1.8259$ and $R_{\rm ej}^{*}=1.8259t_{\rm ini}^{*}$. The velocity and density profiles in characteristic units are given by
\begin{equation}
  v^*(r^*)=\begin{dcases}
    \frac{r^*}{t_{\rm ini}^*}, & \text{$r^*\leq R_{\rm ej}^*$}\\
    0, & \text{$r^*>R_{\rm ej}^*$},
  \end{dcases}
\end{equation}
and
\begin{equation}
  \rho^*(r^*)=\begin{dcases}
    \frac{0.0392}{t_{\rm ini}^{*3}}, & \text{$r^*\leq R_{\rm ej}^*$}\\
    1, & \text{$r^*>R_{\rm ej}^*$}.
  \end{dcases}
\end{equation}

\begin{figure}
    \centering
    \includegraphics[width=\columnwidth]{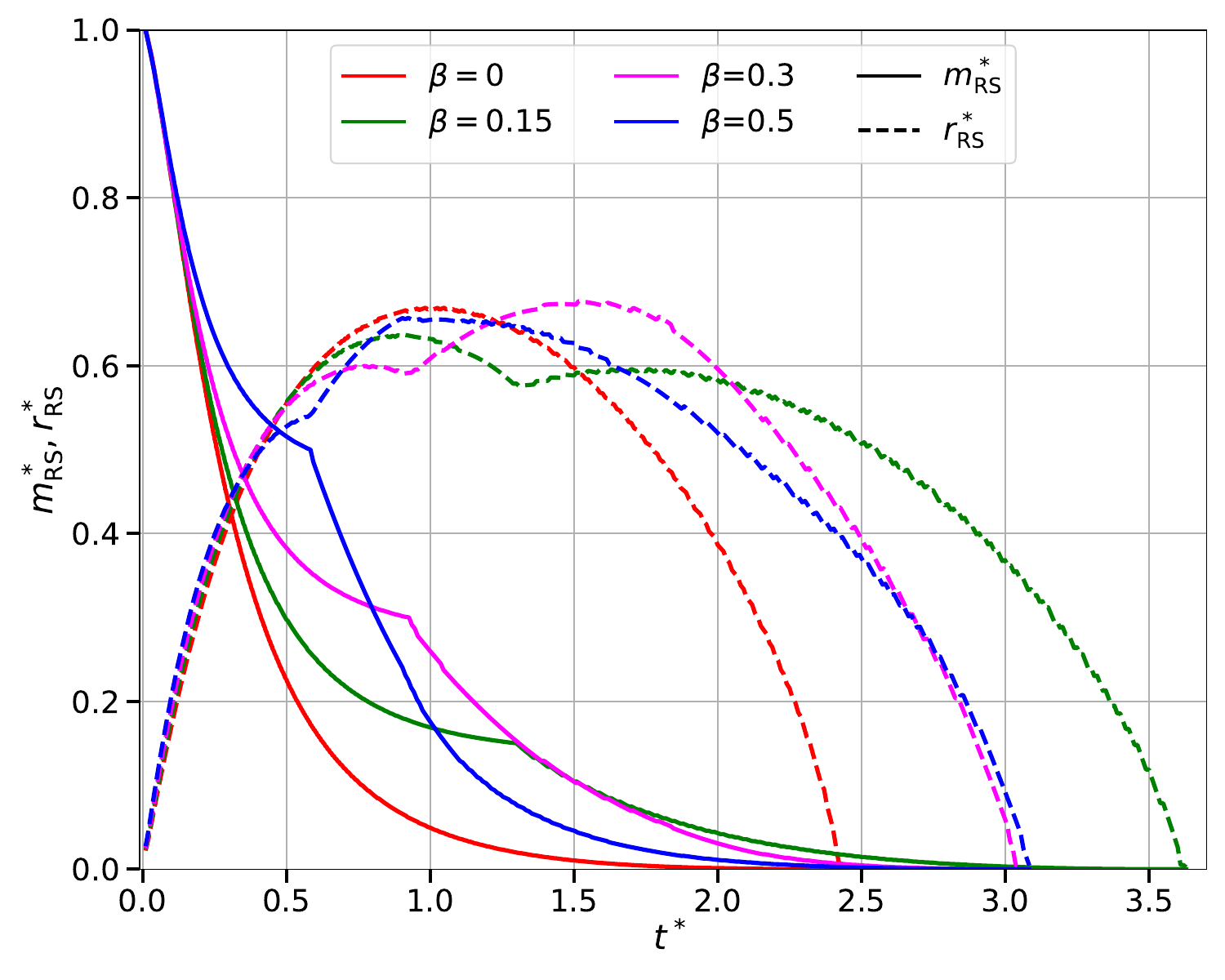}
    \caption{Temporal evolution of the position of the reverse shock front in Eulerian coordinate $r_{\rm RS}$ (dashed lines) and Lagrangian coordinate $m_{\rm RS}$ (solid lines) for the different density profiles in characteristic coordinates. The red, green, magenta, and blue lines show the variation for $\beta=0,0.15,0.3$ and $0.5$ respectively.}
    \label{fig:reverse_shock}
\end{figure}
\subsubsection{Density profiles with a core}
We next consider more realistic density profiles that have a core. We adopt cores of constant density $\rho_{\rm c}$ of mass $M_{\rm c}$ surrounded by a lower density envelope of constant density $\rho_{\rm env}$. The ratio $\beta=M_{\rm c}/M_{\rm ej}$ and $\eta=\rho_{\rm c}/\rho_{\rm env}$ are parameters that are varied depending on the progenitor profile. The profile without any core discussed above corresponds to $\beta=0$ and $\eta=1$. This parametrization of the density profile is roughly consistent with detailed density profiles from the CCSN ejecta from \textsc{kepler} following free expansion. The values of $\beta$ and $\eta$ from the CCSN models range from $\sim 0.14$--$0.6$ and $\sim 10$--$20$, respectively.  Although this parametrization of the density profile introduces two new parameters, because it does not introduce any new length scales, it can still be written in characteristic units to yield \emph{unified} solutions for fixed values of $\beta$ and $\eta$. The density profile is given by
\begin{equation}
  \rho(r)=\begin{dcases}
    \frac{3\kappa_1}{4\pi}\frac{M_{\rm ej}}{R_{\rm ej}^3}, & \text{$r\leq R_{\rm c}$}\\
    \frac{3\kappa_1}{4\pi\eta}\frac{M_{\rm ej}}{R_{\rm ej}^3}, & \text{$R_{\rm c}<r\leq R_{\rm ej}$}\\
    \rho_0, & \text{$r>R_{\rm ej}$},
  \end{dcases}
  \label{eq:rho_core}
\end{equation}
where $\kappa_1=\eta(1-\beta)+\beta$. The radius of the core $R_{c}$ is related to $R_{\rm ej}$ by 
\begin{equation}
    R_{\rm c}= \left( \frac{\beta}{\kappa_1}\right)^{1/3}R_{\rm ej}\equiv\kappa_2R_{\rm ej}
\end{equation}
The above expression for density is very similar to the expression for a profile without a core in Eq.~\ref{eq:rho_nocore}, but with the additional factors of $\kappa_1$, $\eta$, and $\kappa_2$.  
In this case, $R_{\rm ej}$ for a given $t_{\rm ini}$ depends on $\beta$ and $\eta$  in addition to $M_{\rm ej}$ and $E$. As before, $R_{\rm ej}$ can be calculated for a given value of $t_{\rm ini}$ from $v_{\rm ej}$ which is given by 
\begin{equation}
    v_{\rm ej}=\left(\frac{10}{3}\right)^{1/2}\kappa_3E^{1/2}M_{\rm ej}^{-1/2},
\end{equation}
where 
\begin{equation}
    \kappa_3 = \left(\frac{\eta}{\kappa_1}\frac{1}{(\eta-1)\kappa_2^5 +1 }\right)^{1/2}.
\end{equation}
The expression for $v_{\rm ej}$ above is almost identical to the expression for the profile with no core given in Eq.~\ref{eq:vej_nocore} but with the additional factor of $\kappa_3$ which depends on  $\beta$ and $\eta$.
The density profile for a fixed $\beta$ and $\eta$ can again be written in characteristic coordinates with the solution being independent of $E$ and $M_{\rm ej}$. As before, we adopt $t_{\rm ini}^*=0.01$. We explore values of $\beta=0.15, 0.30,$ and $0.50$ that cover the range of mass models. We adopt a value of $\eta=15$ that is roughly consistent with all the models.   

\begin{figure}
    \centering
    \includegraphics[width=1\columnwidth]{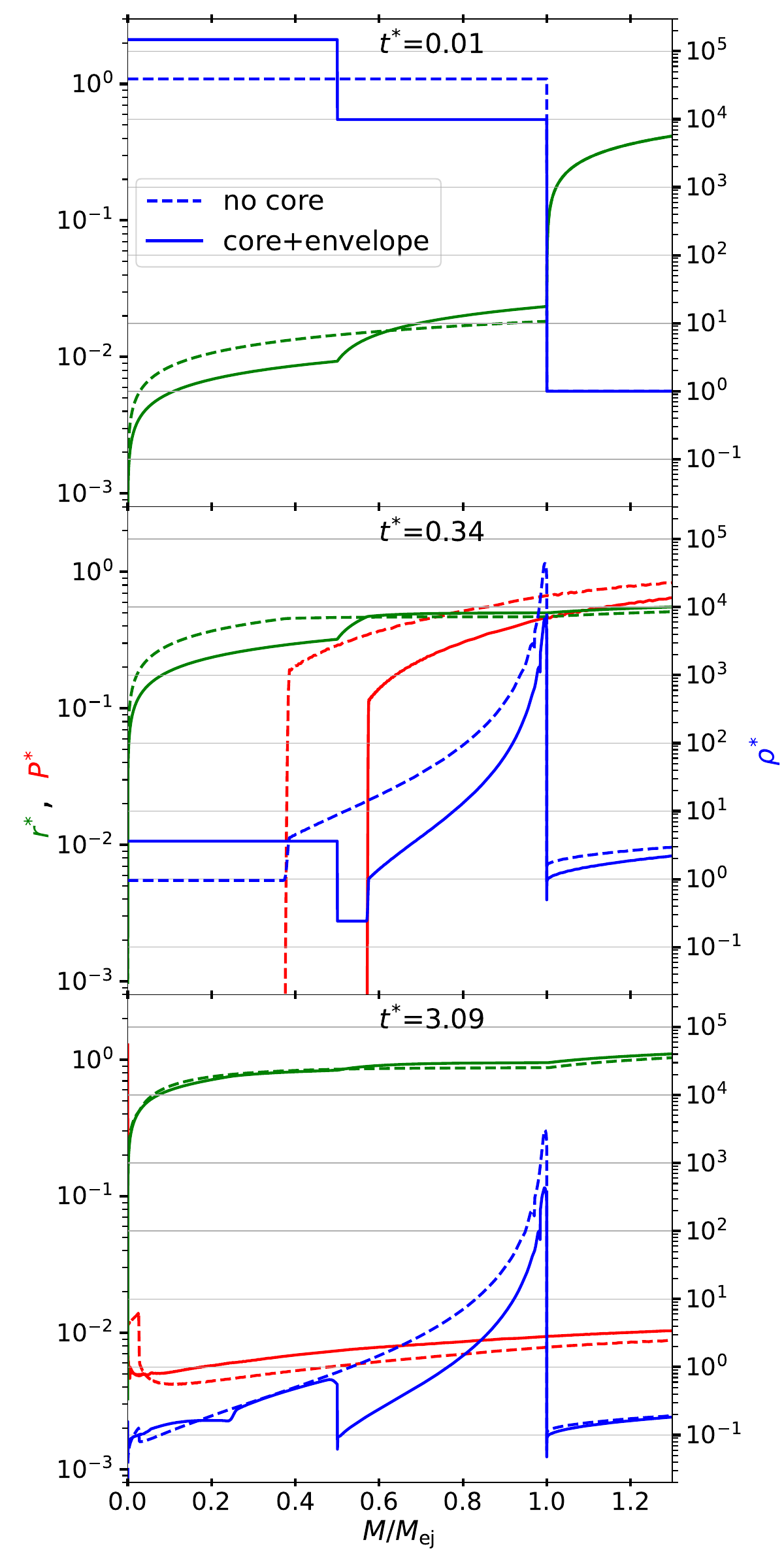}
    \caption{Pressure, mass density, and radius profile as a function of $M/M_{\rm ej}$ at three epochs in the characteristic unit. The solid and dashed line shows the case with {\it core+envelope} ($\beta=0.5$) and {\it no core} respectively. The blue color shows the mass density profile (right y-axis). The red and green color shows the pressure and radius profile (left y-axis) respectively. The {\it top panel} shows the initial setup at $t^* = 0.01$. The {\it middle panel} shows the profiles at $t^* = 0.34$ when the reverse shock is about to hit the core for {\it core+envelope} case. The {\it bottom panel} shows the profiles when the reverse shock has hit the center of the ejecta at $t^* = 3.09$ for {\it core+envelope} case. For {\it no core} case, the reverse shock has already hit the center by $t^* = 3.09$ and the secondary shock wave has started traveling outwards as can be seen by the jump in pressure (dashed red line) and density profile (dashed blue line).}
    \label{fig:prs-profile}
\end{figure}
\begin{figure}
    \centering
    \includegraphics[width=1.0\columnwidth]{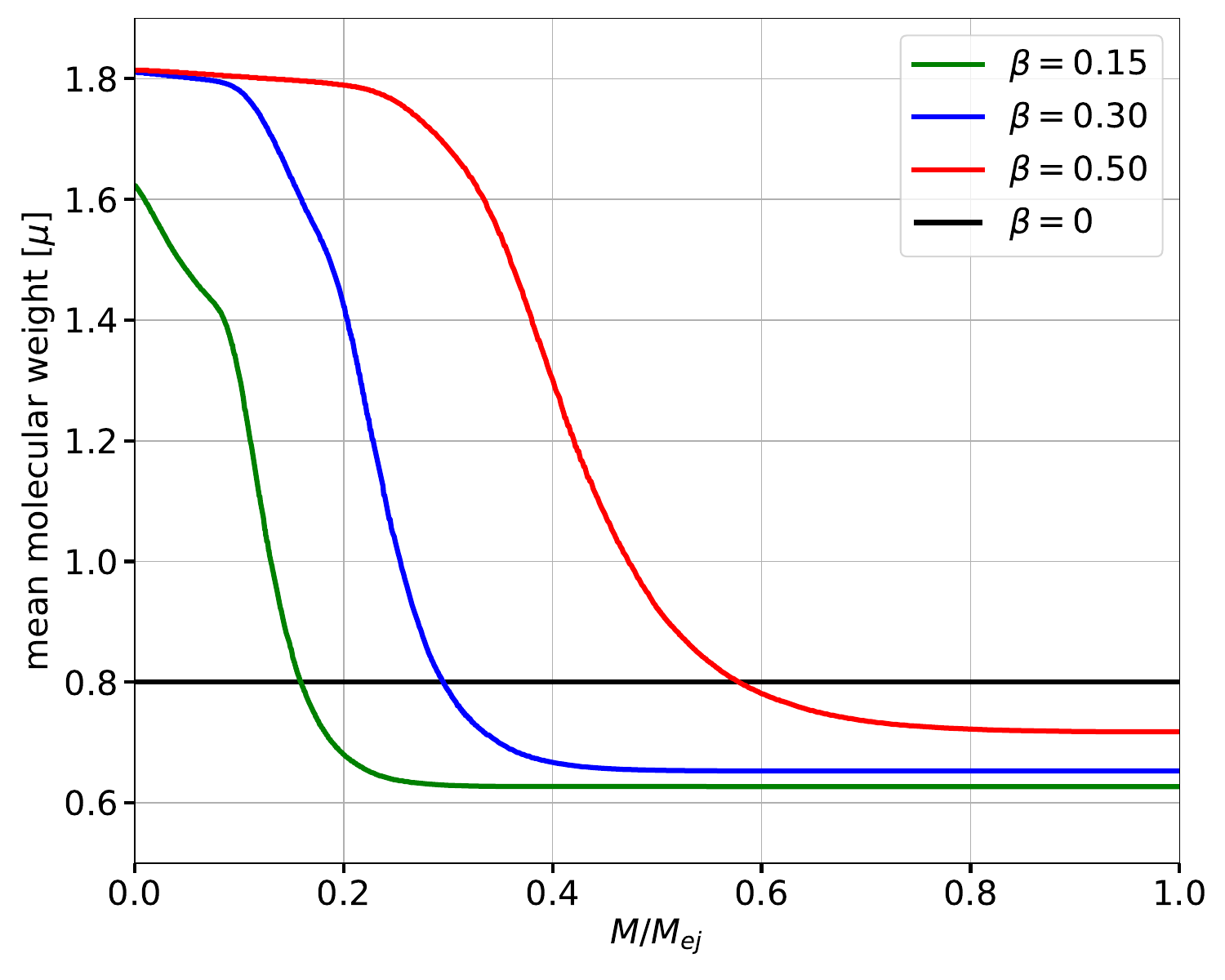}
    \caption{Mean molecular weight profile as a function of $M/M_{\rm ej}$ calculated using \textsc{kepler}. The black, green, blue, and red lines show the mean molecular weight profile for $\beta=0,0.15,0.30$, and $0.50$ respectively.}
    \label{Fig:mu_var}
\end{figure}
\begin{figure}
    \centering
    \includegraphics[width=1.0\columnwidth]{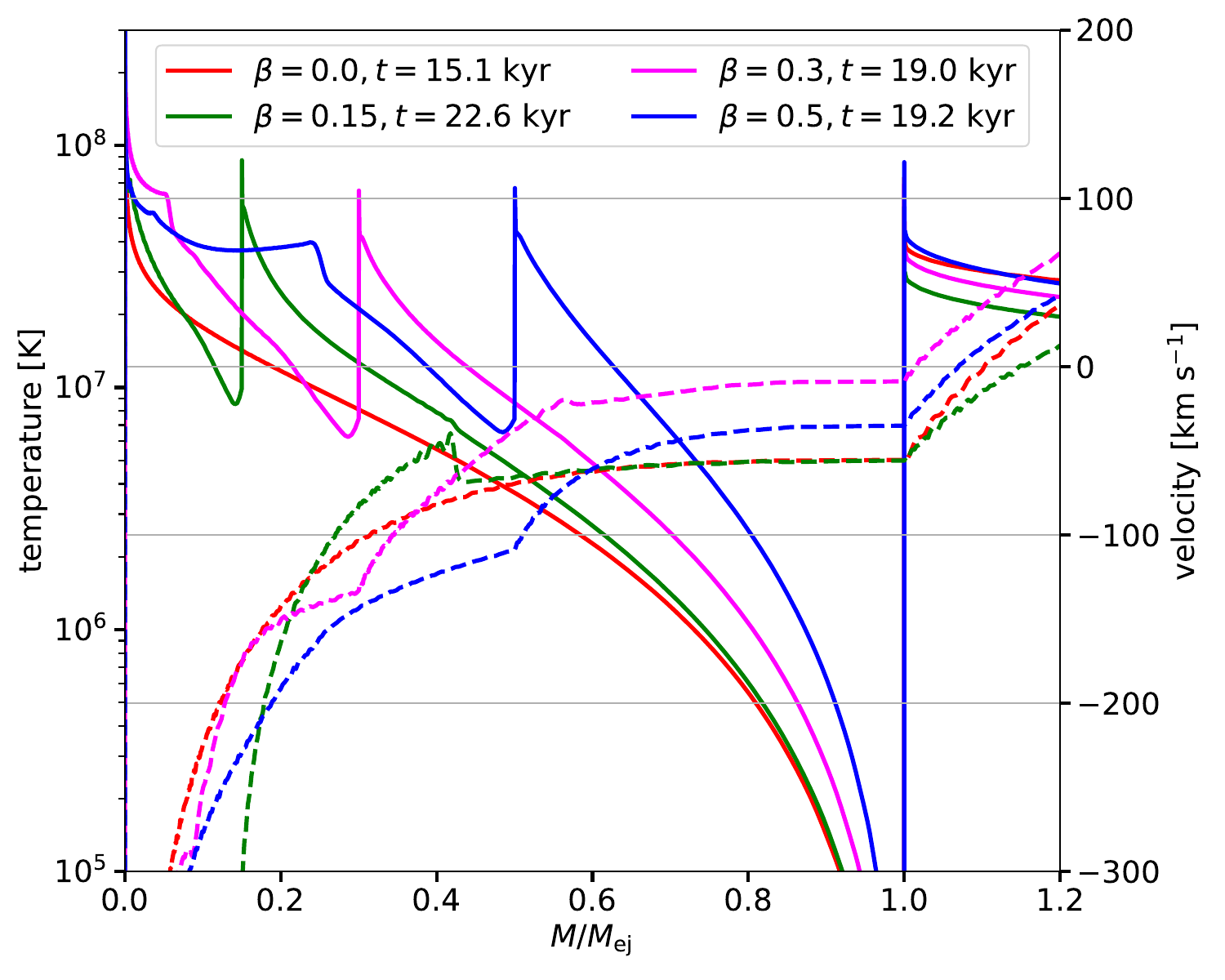}
    \caption{Temperature (solid lines) and velocity profiles (dashed lines) for different values of $\beta$ when the reverse shock hits the center of the ejecta in the respective cases. Red, green, magenta, and blue colors show the cases with $\beta = 0, 0.15, 0.3$, and $0.5$ respectively. The profiles are plotted for the fiducial case with $E=10^{51}\,{\rm erg}$, $M_{\rm ej}=10 \ M_{\odot}$ and $n_0=0.1\,{\rm cm^{-3}}$.}
    \label{Fig:temp-profile}
\end{figure}

\section{Results}\label{sec:results}
\subsection{Overall Evolution of the Ejecta}
As described in \citep{Truelove1999}, because the freely expanding ejecta has velocity that is orders of magnitude larger than the sound speed of the ambient medium, a forward blast wave is launched into the ambient medium that leads to compression and heating. This in turn exerts pressure back on the freely expanding ejecta that compresses ejecta and launches a reverse shock. The shocked ejecta (due to reverse shock) is always separated from the less dense shocked ambient medium by the contact discontinuity at $M=M_{\rm ej}$. The reverse shock continues to travel inwards (in mass coordinates) and results in the heating up of the shocked ejecta due to compression. Because the velocity of the initial freely expanding ejecta decreases with radius, the reverse shock progressively encounters slower ejecta velocities as it moves inward. 
The reverse shock eventually reverses the direction of the expanding ejecta leading to an implosion at $t^{*}\gtrsim 1$.  This can be seen clearly from Fig.~\ref{fig:reverse_shock}, where the dashed and solid red lines show the evolution of the position of the reverse shock in Eulerian coordinates ($r_{\rm RS}$) and Lagrangian coordinates ($m_{\rm RS}$), respectively,  for the density profile without a core. We note here that the location of the reverse shock front is defined as the coordinate where the pressure jumps from zero (corresponding to unshocked ejecta) to a non-zero value (corresponding to shocked ejecta). By this time the forward blast wave in the ambient medium transitions to the ST phase. Eventually, the reverse shock reaches the center at $t^{*}\gtrsim 2$ leading to an outward-going secondary blast-wave shock referred to as the \emph{bounce}. As mentioned in \citet{Truelove1999}, the secondary blast-wave shock reverses the direction of the infalling material that was previously shocked by the reverse shock. 

The propagation of reverse shock discussed above applies only to density profiles without a core. In the case of the density profiles with a core, when the reverse shock reaches the core--envelope boundary, the large density gradient causes a bounce and launches a forward secondary blast-wave shock wave that travels outwards. The corresponding position of the reverse shock for the various values of $\beta$ are shown in Fig.~\ref{fig:reverse_shock}. As can be seen from the figure, the reverse shock is heavily slowed down by the bounce at the core--envelope boundary, and starts to move out again in Eulerian coordinates (dashed green, magenta, and blue lines) even though it continues to move inwards in Lagrangian coordinates (solid green, magenta, and blue lines).  
In the case of density profiles with cores, the reverse takes longer to reach the center with profiles with the smallest core taking the longest (see Fig.~\ref{fig:reverse_shock}). When the secondary blast-wave shock reaches the contact discontinuity another secondary reverse shock is launched and the process repeats itself. During this, the original reverse shock in the core eventually reaches the center leading to bounce and is helped along by subsequent arrivals of secondary reverse shocks.

In all cases, we end our simulation at $t^{*}=10$, well after the reverse shock has reached the center. Following bounce when reverse shock reaches the center, 3D effects such as inhomogeneities and turbulent mixing due to Raleigh-Taylor instabilities become increasingly important at the contact discontinuity with multiple cycles of secondary shock waves resulting in the contact discontinuity being erased and the flow subsequently evolves as a ST solution throughout the remnant \citep{Truelove1999}.

\subsection{Evolution of Pressure, Density, and Temperature of Shocked Ejecta}

In order to illustrate the evolution of the structure of the ejecta we consider the density profiles with $\beta$ of 0 and 0.5. Figure \ref{fig:prs-profile} shows the pressure and density profiles in characteristic coordinates as a function of the mass coordinate at various instants of the evolution ranging from the initial to the final stage when the reverse shock reaches the center. The radial (Eulerian) coordinate $r$ is also plotted as a function of the mass coordinate. The evolution of pressure and density for both profiles are largely similar where the compression due to the reverse shock leads to a pressurized ejecta with a relatively flat pressure profile in terms of the mass coordinate. 
The density of the shocked ejecta is highest at the contact discontinuity at $m^{*}=1$ and decreases with decreasing $m^{*}$ as the reverse shock moves inwards.

By the time the reverse shock has reached the center, the central density is lower by $\sim$ four orders of magnitude compared to the highest density at the contact discontinuity. 
By this time, all the matter in the ejecta is compressed into a thin shell in Eulerian coordinates and almost all of the ejecta is confined to a narrow range of radial coordinates. In profiles with a core, the jump in density due to the core and the associated slowdown of the reverse shock discussed earlier leads to a shallower density gradient in the core by the time the reverse shock has reached the center.

The temperature $T$ can be  calculated from the pressure and density assuming an ideal gas law given by
\begin{equation}
    T=\frac{\mu m_u}{ k_{\rm B}}\frac{P^{*} P_{\rm ch}}{\rho^{*} \rho_{\rm ch}},
\end{equation}
where $\mu$ is the mean molecular weight of the ejecta, $m_u$ is the atomic mass unit, and $k_{\rm B}$ is the Boltzmann constant. Because of the metallicity gradient with an increasing amount of metals in the core, for a given progenitor, $\mu$ increases with decreasing mass. Figure~\ref{Fig:mu_var} shows the variation of $\mu$ with mass coordinate for a $11, 18,$ and $25\,\Msun$ models corresponding to $\beta$ of $0.15, 0.3,$ and $0.5$, respectively, assuming a fully ionised ejecta. For the profile without a core i.e., $\beta=0$, we adopt a constant $\mu=0.8$ which corresponds to the mass averaged value for a typical CCSN progenitor over the ejecta mass. Because the unified solution in characteristic coordinates for a given density and $\mu$ profile remains unchanged, $T\propto P_{\rm ch}/\rho_{\rm ch}=E M_{\rm ej}^{-1}$. 
Figure~\ref{Fig:temp-profile} shows the temperature and velocity for $E=10^{51}\,{\rm erg}$, $M_{\rm ej}=10\,\Msun$, and $n_0=0.1\,{\rm cm^{-1}}$ when the reverse shock has reached the center for the various density profiles. As can be seen from the figure, for density profiles with a core, the temperature (solid green, magenta, and blue lines) of the shocked ejecta in the core is $\gtrsim 6\times 10^6\,{\rm K}$ in all cases. For the density profile without a core (solid red line), the inner $\sim 40\%$ of the inner ejecta lies above this temperature. The maximum temperature occurs at the center due to a combination of lowest density and highest $\mu$ with a maximum temperature of $\sim 6\times 10^{7}\,{\rm K}$. Because $T\propto E$, for a given density profile, increasing $E$ by some factor will lead to an overall increase in $T$ by the same factor. For example, for $E=10^{52}\,{\rm erg}$, $T$ will increase by a factor of 10 compared to what is plotted in Fig.~\ref{Fig:temp-profile}. 
The corresponding velocity profiles for the various values of $\beta$ are also shown in Fig.~\ref{Fig:temp-profile} (dashed lines). As mentioned earlier, by the time the reverse shock reaches the center, the explosion turns into an implosion and the velocity of the entire ejecta is negative with values of $\lesssim 0$ and $\lesssim -300\,{\rm km\,s^{-1}}$ at the contact discontinuity and the center, respectively. 

\begin{figure}
    \centering
    \includegraphics[width=1.0\columnwidth]{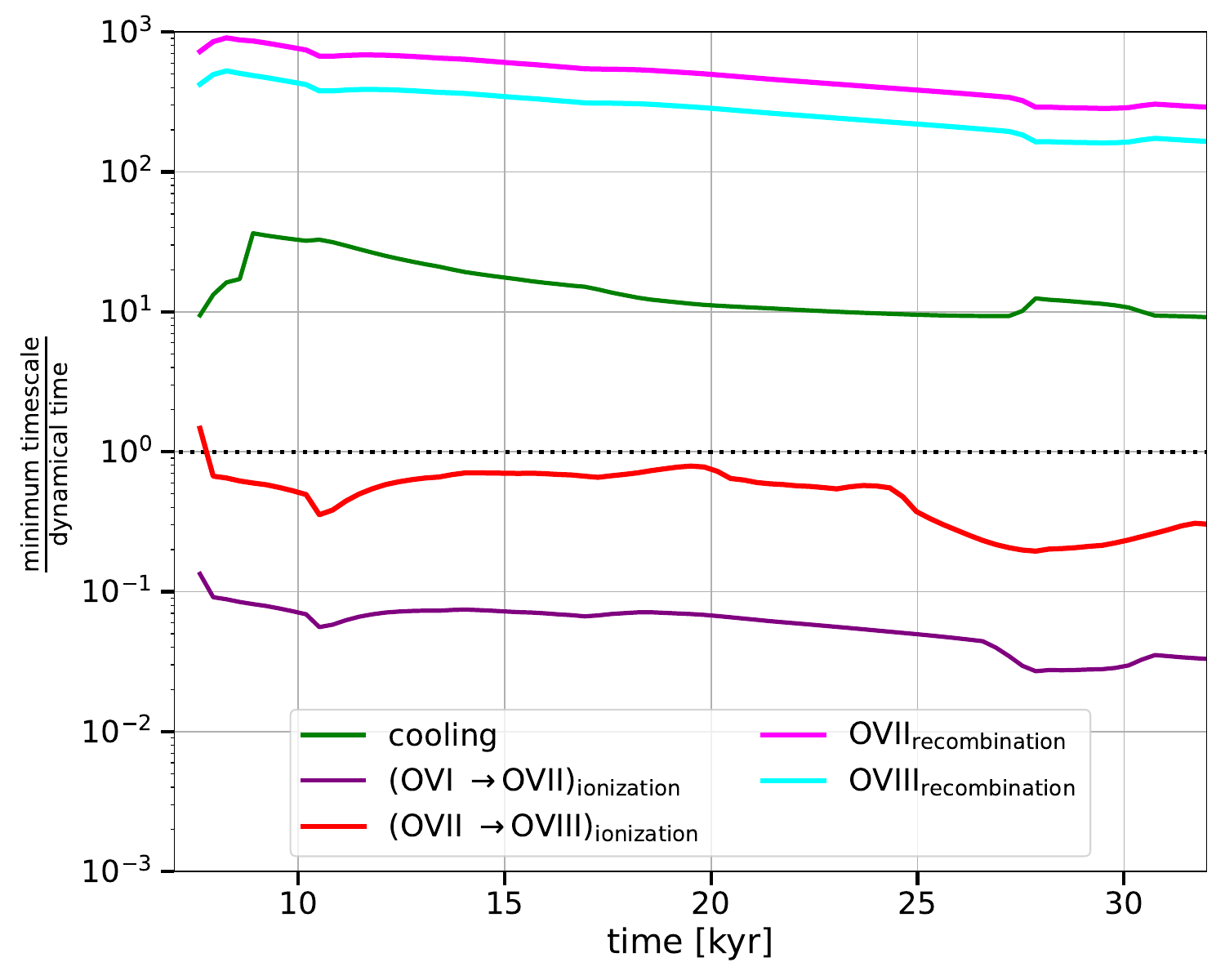}
    \caption{This figure shows the ratio of "minimum" radiative cooling, recombination, and ionization timescale to the dynamical time as a function of time. The "minimum" timescale corresponds to the mass shell which has minimum radiative cooling/ionization/recombination time within $M/M_{\rm ej} \leq 0.2$ and $T\geq 4\times 10^6$ K. The green line shows the ratio of radiative cooling to dynamical time calculated using the cooling function for core composition with \textsc{cloudy} (\citet{Ferland2017}). The magenta and cyan lines show the ratio of the recombination timescale (\citet{Verner1996}) to the dynamical timescale for OVII and OVIII respectively. The purple and red lines show the ratio of the ionization timescale (\citet{Voronov1997}) to the dynamical time for OVI and OVII respectively. The above timescales are plotted for $E=10^{51}$ erg, $M_{\rm ej}=10 \ M_{\odot}$, $n_0=0.1 \rm cm^{-3}$ and $\beta=0.3$.}
    \label{fig:timescales}
\end{figure}

\subsection{Comparing Timescales}\label{sec:timesclaes}
Before we discuss column densities, it is important to consider the ionization, recombination, and cooling timescales of the shocked heated ejecta relative to the dynamical time. The radiative cooling time for the gas in the ejecta is $t_{\rm c} = 3k_{B}T/(2n_{\rm ion}\Lambda)$, where $\Lambda$ (erg cm$^3$ s$^{-1}$) is the cooling function and $n_{\rm ion}$ is the total number density of ions. The recombination timescale for an ion is $t_{\rm re} = 1/(n_e\alpha_{\rm ion}(T))$, where $\alpha_{\rm ion}(T)$ is the recombination coefficient of the ion at temperature T (\citet{Verner1996}) and $n_e$ is the number density of electrons. The ionization timescale for an ion due to thermal electrons is $t_{\rm ion}=1/(n_e q_i(T))$ where $q_i$ is the ionization coefficient of the ion (\citet{Voronov1997}). Figure~\ref{fig:timescales} shows the minimum ionization and recombination timescales for OVI/OVII and OVII/OVIII along with the cooling timescale relative to the dynamical timescale of the material in the inner core of the ejecta with $\beta=0.3$ that contains almost all of the metals. As can be seen from the figure, the fastest cooling time is an order of magnitude longer than the dynamical time implying that cooling is essentially negligible on dynamical timescales. This justifies our assumption of neglecting the cooling of the shocked ejecta.  

The ionization timescale on the other hand, is shorter than the dynamical timescale, whereas the recombination timescales are more than two orders of magnitude longer. Thus, even though ionization can occur as the reverse shock heats up the ejecta, recombination is essentially negligible. Consequently, CIE is not applicable for the reverse shock heated ejecta and cannot be applied for estimating the abundance of ions. On the other hand, because recombination is negligible, the time evolution of the ionization of the ejecta can be easily calculated which can then be used to calculate the column density of different ionization states for an element and compared directly with the observed values (Table \ref{tab:data}).
However, in order to calculate the column densities of different ionization states, we need to first calculate the total column density of the corresponding element. We discuss this in detail in the following section.

\begin{figure}
    \centering
    \includegraphics[width=1\columnwidth]{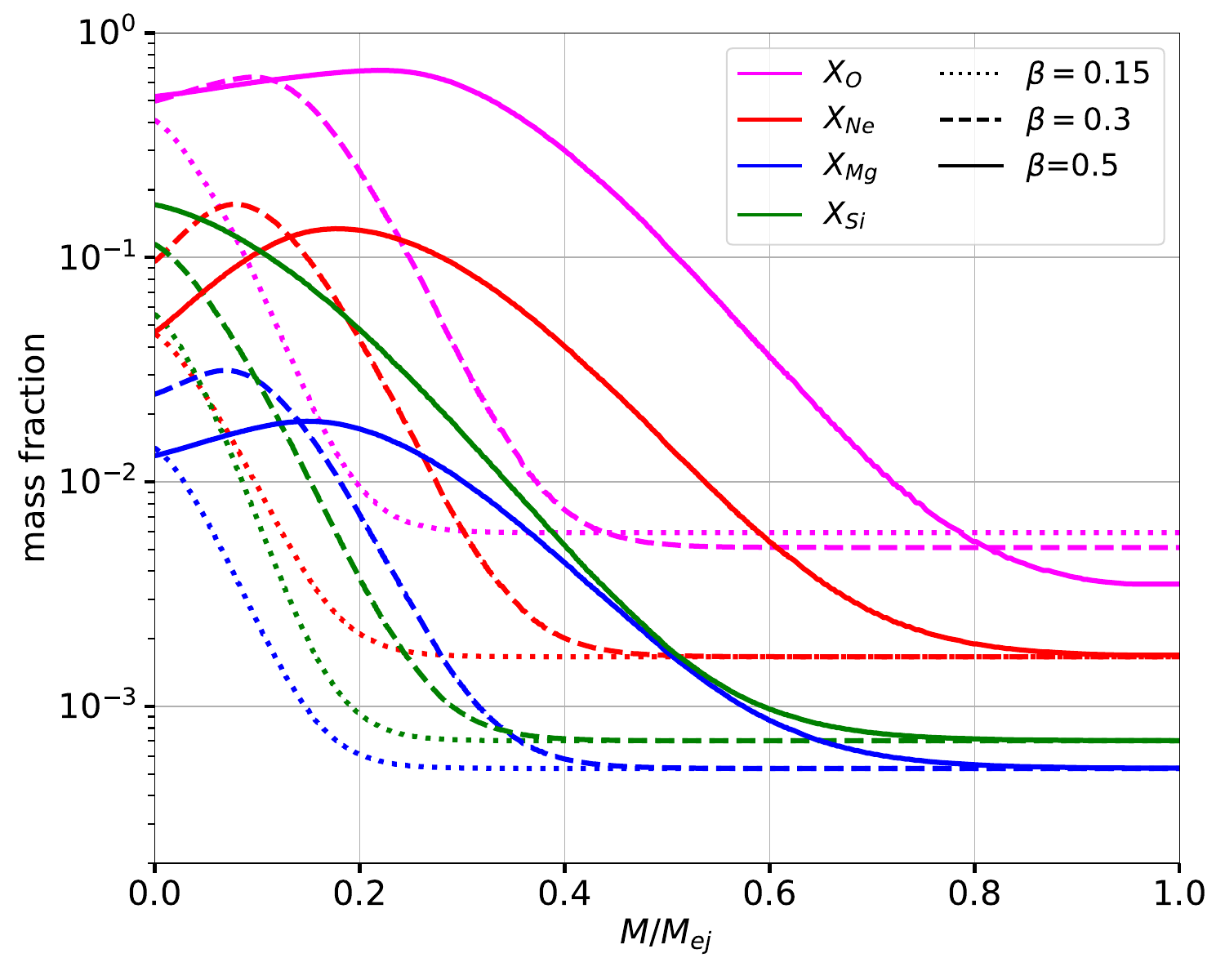}
    \caption{Mass fraction of different elements as a function of $M/M_{\rm ej}$ obtained using \textsc{kepler}. The magenta, red, blue, and green lines show Oxygen, Neon, Magnesium, and Silicon mass fraction profiles respectively. The solid, dashed, and dotted lines refer to the case with $\beta=0.5,0.3$ and $0.15$  corresponding to $25$, $18$, and $11$ M$_{\odot}$ mass models, respectively.}
    \label{fig:mass_frac_var}
\end{figure}

\begin{figure}
    \centering
    \includegraphics[width=1\columnwidth]{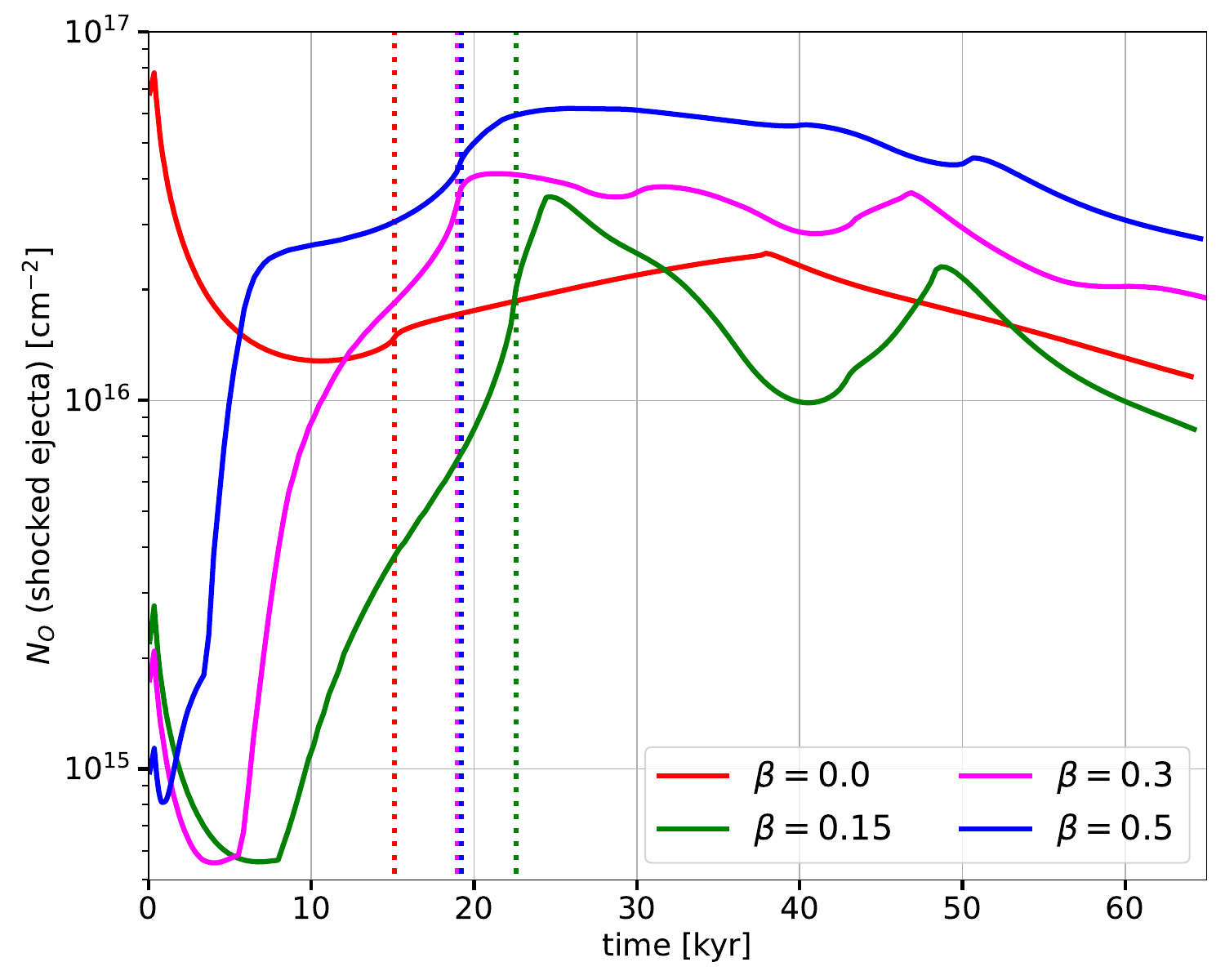}
    \caption{Evolution of Oxygen column density of the shocked ejecta with time for density profiles with different core sizes of $\beta=0, 0.15, 0.3$, and $0.5$ shown in red, green, magenta, and blue, respectively. For {\it no core} case ($\beta=0$), we use a constant Oxygen mass fraction of $0.15$ throughout the ejecta. The dotted vertical lines in the respective color show the time when the reverse shock hits the center of the ejecta. The curves correspond to the adopted fiducial values of $E=10^{51}\,{\rm erg}$, $M_{\rm ej}=10\,\Msun$, and $n_0=0.1\,{\rm cm^{-3}}$.}
    \label{fig:col_beta_var}
\end{figure}

\subsection{Evolution of Column Density}
We calculate the time evolution of the total elemental column densities across all ionization states for Oxygen, Neon, and Silicon in the shock-heated ejecta. The column density $N$ at time $t$ for any element in the shocked ejecta is given by
\begin{equation}
    N(t)= 2\int_0^{R_{\rm ej}(t)} \frac{X \rho (t)}{m_u A}dr,
    \label{eq:column_den}
\end{equation}
where $X$ is the mass fraction of the element and $A$ is the mass number. The factor of 2 is to account for the fact that light from a background source traverses the material in the shell formed by the SNR twice before reaching Earth. We note here that for realistic density profiles i.e., with cores, almost the entire contribution to column density for a metal is from the shocked core with negligible contribution from the envelope. We note here that because the shocked core has temperatures $\gg 10^6\,{\rm K}$  for all the density profiles as discussed in the previous section, the contribution to the column density is essentially from the SV phase. 

Since the characteristic solution remains unchanged for a given density profile and $N$ has the dimensions of number density multiplied with distance, the scaling relation for $N\propto \rho_{\rm ch}R_{\rm ch}=\rho_0^{2/3} M_{\rm ej}^{1/3}$. 
Because $M_{\rm ej}$ varies only slightly from $9$--$13\,\Msun$ across all progenitors with a dependence of $N\propto M_{\rm ej}^{1/3}$, column density is effectively independent of the ejecta mass. The ambient density $\rho_0$ is the only parameter that can potentially have an appreciable effect on $N$ for a given density profile. The nature of the progenitor also affects the column density primarily due to the size of the core and the slightly different metal fractions in the core. Figure~\ref{fig:mass_frac_var} shows the elemental mass fraction profile for the progenitors with different core sizes corresponding to different progenitor masses. The mass fraction of metals is slightly higher in more massive models that have larger cores. This leads to an overall enhancement of the column density of metal as it directly scales with the mass fraction (see Eq.~\ref{eq:column_den} above). 

To illustrate the evolution of column density with time for various progenitors, we adopt the fiducial $E=10^{51}\,{\rm erg}$, $M_{\rm ej}=10\,\Msun$, and $n_0=0.1\,{\rm cm^{-3}}$ corresponding to $t_{\rm ch}=6.2\,{\rm kyr}$. Figure~\ref{fig:col_beta_var} shows the evolution of the Oxygen column density of the shocked ejecta for density profiles with different core sizes. 
As can be seen from the figure, the column density starts with a relatively high value due to the high initial density near the contact discontinuity but starts to decrease as the shocked ejecta keeps expanding.
The column density eventually increases with time as the reverse shock moves inward and more metal is heated up. The increase is particularly dramatic for density profiles with cores where the column density starts to rise rapidly when the reverse shock reaches the core which takes longer for profiles with smaller cores.  
The instant when the reverse shock hits the center is shown using dotted vertical color lines in figure \ref{fig:col_beta_var} for various values of $\beta$. As mentioned earlier, by this time the ejecta to temperatures of $\gtrsim 6\times 10^6\,{\rm K}$ corresponding to `super-virial' temperatures. Even after the shock reaches the center, the column density increases slightly due to increased density due to implosion. Following the bounce at the center, the column density effectively plateaus out and remains roughly constant until the end of the simulation. This is primarily because almost all of the ejecta is confined to a narrow range of radius by the time the shock reaches the center. Consequently, to zeroth order, all the material in the ejecta is roughly at the same radius $R_{\rm c}$. This implies that 
\begin{equation}
    N\propto \int_0^{R_{\rm ej}}\rho dr=\int_0^{R_{\rm ej}} \frac{dm}{4\pi r^2}\approx  \frac{M_{\rm c}}{4\pi R_{\rm c}^2}.
\end{equation}
Following the bounce at the center, the implosion is gradually turned into an explosion after multiple secondary blast waves are generated as described earlier. The average velocity of the shocked ejecta $\bar v_{\rm ej}$, however, remains very low including being negative during some intervals during a series of implosions and explosions. As a result, $R_{\rm c}$ and consequently $N$ changes slowly on timescales of $\gg t_{\rm ch}$. For example, compared to the peak value after the bounce $t^*\sim 3$, the value of  $N$ decreases by a factor of $2.7$, $4.4$, $2.2$, and $2.2$,  for density profiles with $\beta$ of 0, 0.15, 0.3, and 0.5, respectively, by $t^*=10$. It is important to note that after bounce, due to repeated cycles of implosions and explosion, the column density does not decrease monotonically with time as can be seen from Fig.~\ref{fig:col_beta_var}. So, the effective timescale during which the column density changes is even longer than what is estimated above ($t^\ast=10$). 

\subsection{Comparison with Observations}\label{subsec:comp_obs}

\begin{figure}
    \centering
    \includegraphics[width=\columnwidth]{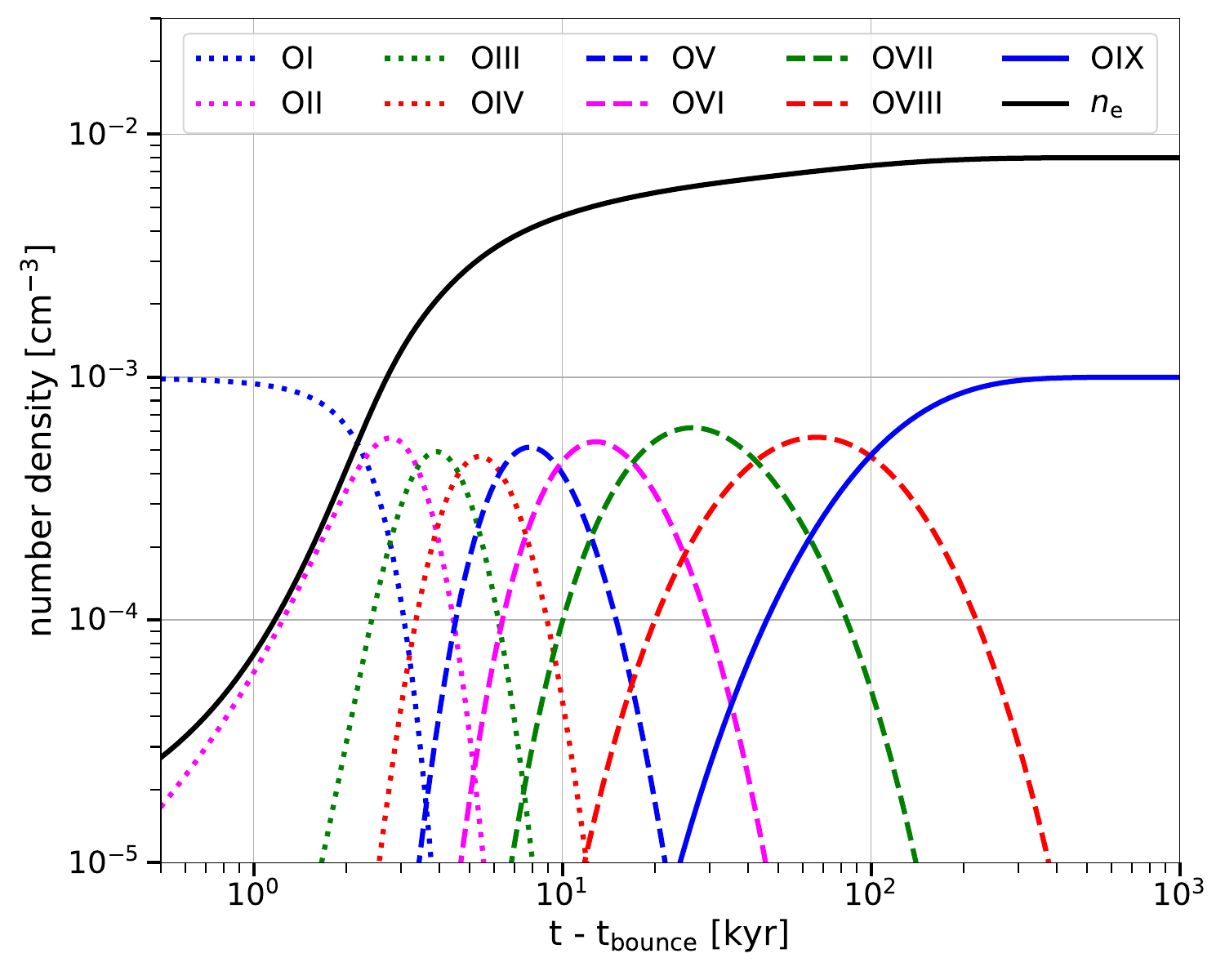}
    \caption{Ionization of a Oxygen gas at $T=5$ keV with initial OI abundance of $0.001\, {\rm cm^{-3}}$ and initial $n_e=10^{-5}\, {\rm cm^{-3}}$. }
    \label{fig:neon_ionization}
\end{figure}
We now compare the theoretical column densities for Oxygen, Silicon, and Neon ions with the observed range of values. Because $N$ is not very sensitive to $E$ and $M_{\rm ej}$, we consider the fiducial values of $10^{51}\,{\rm erg}$ and $10\,\Msun$, respectively. We consider an ambient density of $n_0=0.1\,{\rm cm^{-3}}$ for the three profiles with cores. Here we do not consider the profile without a core as it does not correspond to any physical model. 

In order to compare our results with the observed column density of ions, the ion fraction of OVIII, NeX and SiXIV are required, for which we use the following procedure. Since the recombination timescale is much larger than the dynamical time (cyan and magenta lines in figure \ref{fig:timescales}), the recombination of ions is neglected in our calculation. 
We solve the system of coupled linear differential equations governing the ionization equation to find the abundance of various ionization states of various elements as a function of time starting from the neutral to the fully ionized state. The rate of change of the abundance of the $i^{\rm th}$ state of ionization of an atom depends on the rate of ionization of the $i^{\rm th}$ to $(i+1)^{\rm th}$ state. This is counteracted by the production of the $i^{\rm th}$ state by the ionization of the $(i-1)^{\rm th}$ state. 
We adopt the ionization rates from \citet{Voronov1997} and consider ionization due to thermal electrons. The ionization parameter $q$, depends only on the temperature of the electrons. 

The ionization essentially occurs after the reverse shock heats up all the metal in the ejecta i.e., after bounce. As discussed earlier, the mass density of the shocked ejecta varies very slowly after bounce. Moreover, the cooling timescale is an order of magnitude longer than the dynamical timescale (green line in figure \ref{fig:timescales}). For this reason, for simplicity, we assume that the temperature and mass density remain constant while solving for the ionization. We adopt $n_{\rm ion}\approx 0.001$--$0.002\,{\rm cm^{-3}}$, which is taken from the density profile of the ejecta after the reverse shock has hit the center ($t\sim 20$ kyr). Since the temperature of the shocked core varies somewhat with the density profile, we take a range of $3\hbox{--}10$ keV for Oxygen, Magnesium, and Neon and $10\hbox{--}18$ keV for Silicon. These numbers are adopted directly from the range of temperatures found in the inner part of the ejecta ($M/M_{\rm ej} <0.2$) as can be seen in Fig.~\ref{Fig:temp-profile}.

For illustration, we show the evolution of the ionization for a gas composed of Oxygen with an initial OI abundance of  $0.001 {\rm cm^{-3}}$ and initial $n_e=10^{-5}$ corresponding to an initial ionization fraction of $\sim 0.1\%$. Figure~\ref{fig:neon_ionization} shows the evolution of OI--OIX along with $n_e$, for electron temperature of 5 keV. As can be seen from the figure, $n_e$ increases rapidly on timescales of $\sim$ kyr and reaches an ionization fraction of $\sim 80\%$ within 40 kyr as Oxygen gets progressively more ionized. Beyond this epoch the gas is primarily composed of OVII--OIX and full ionization is reached by $\sim 200\,{\rm kyr}$.

\begin{figure}
    \centering
    \includegraphics[width=1\columnwidth]{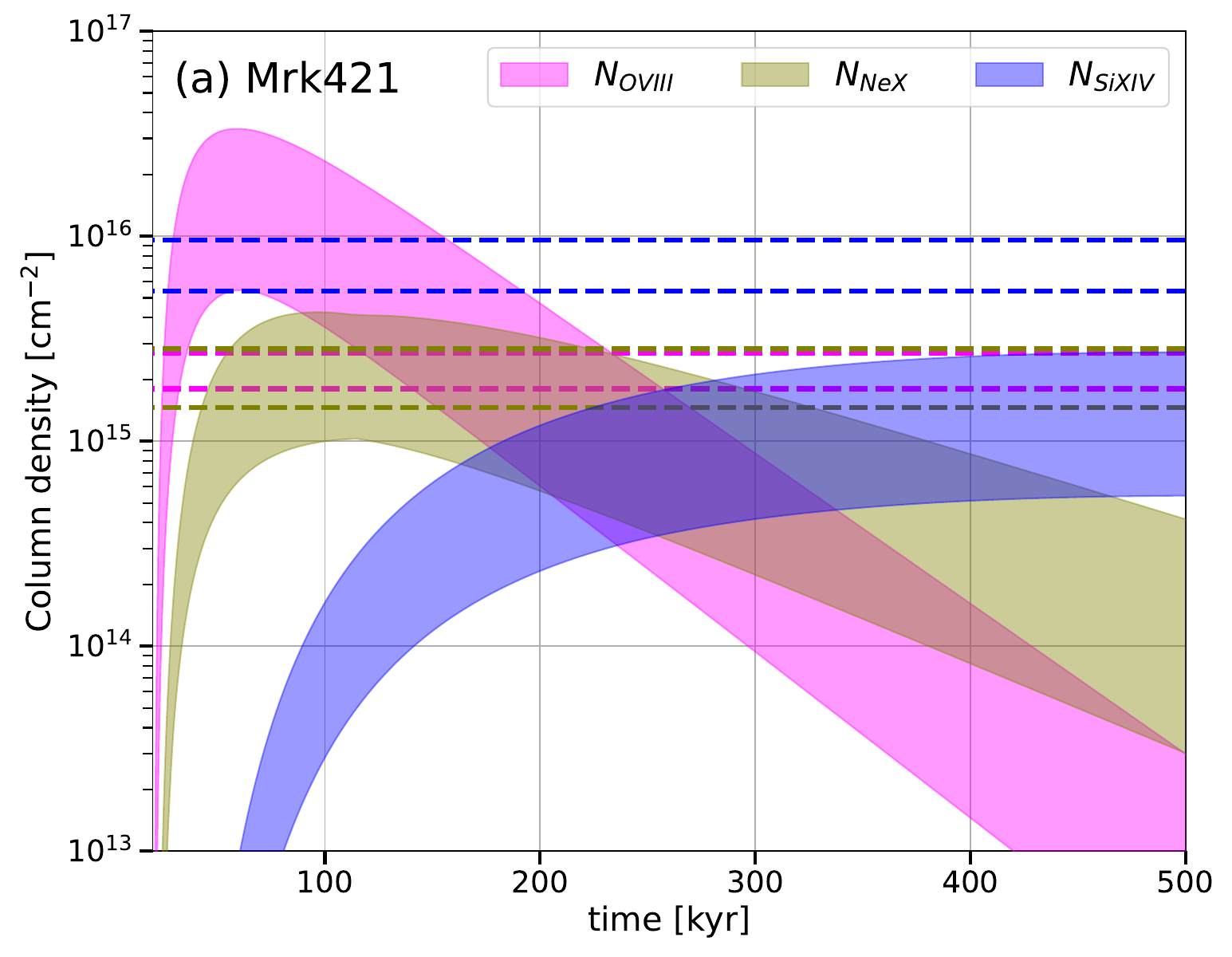}
    \includegraphics[width=1\columnwidth]{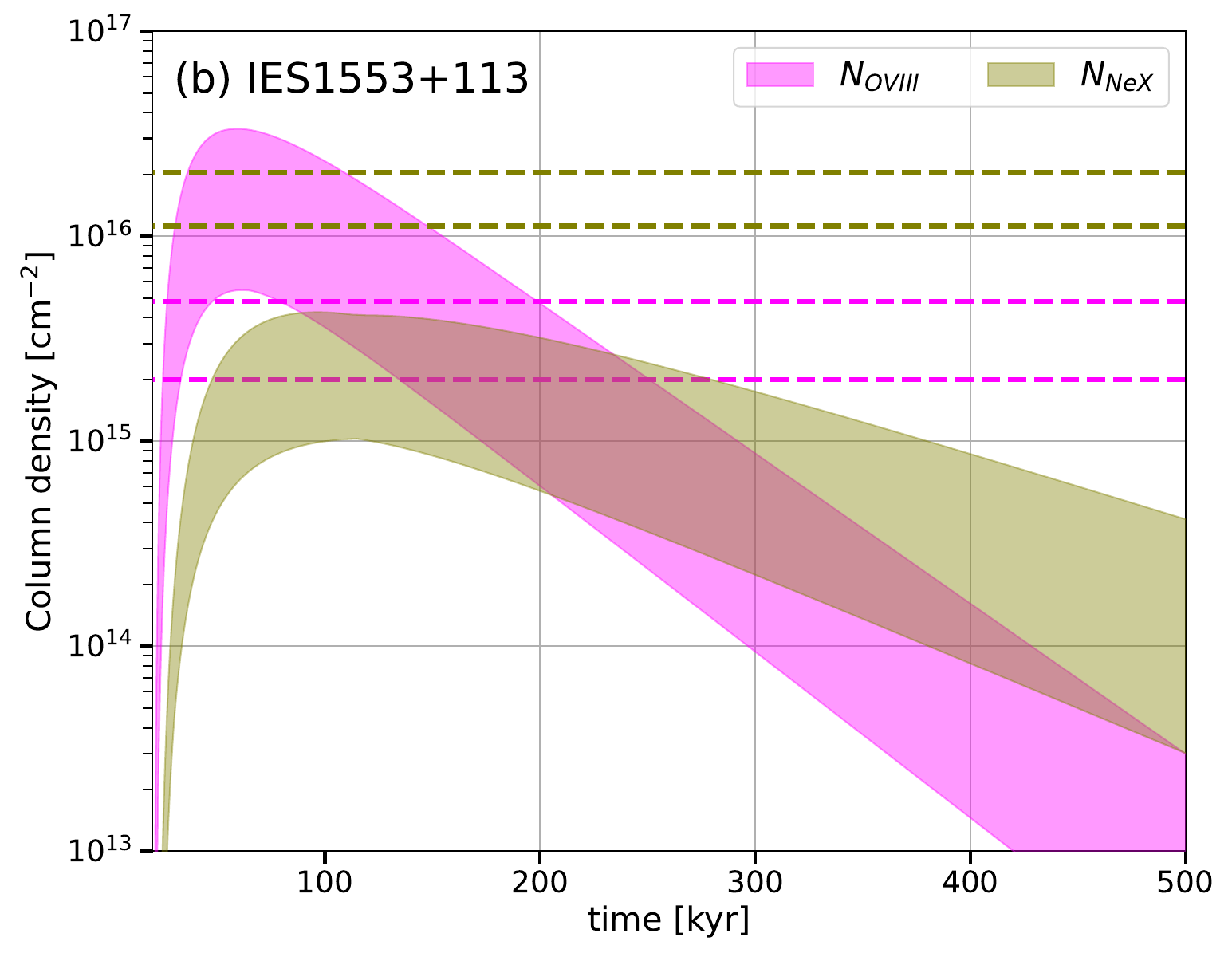}
    \includegraphics[width=1\columnwidth]{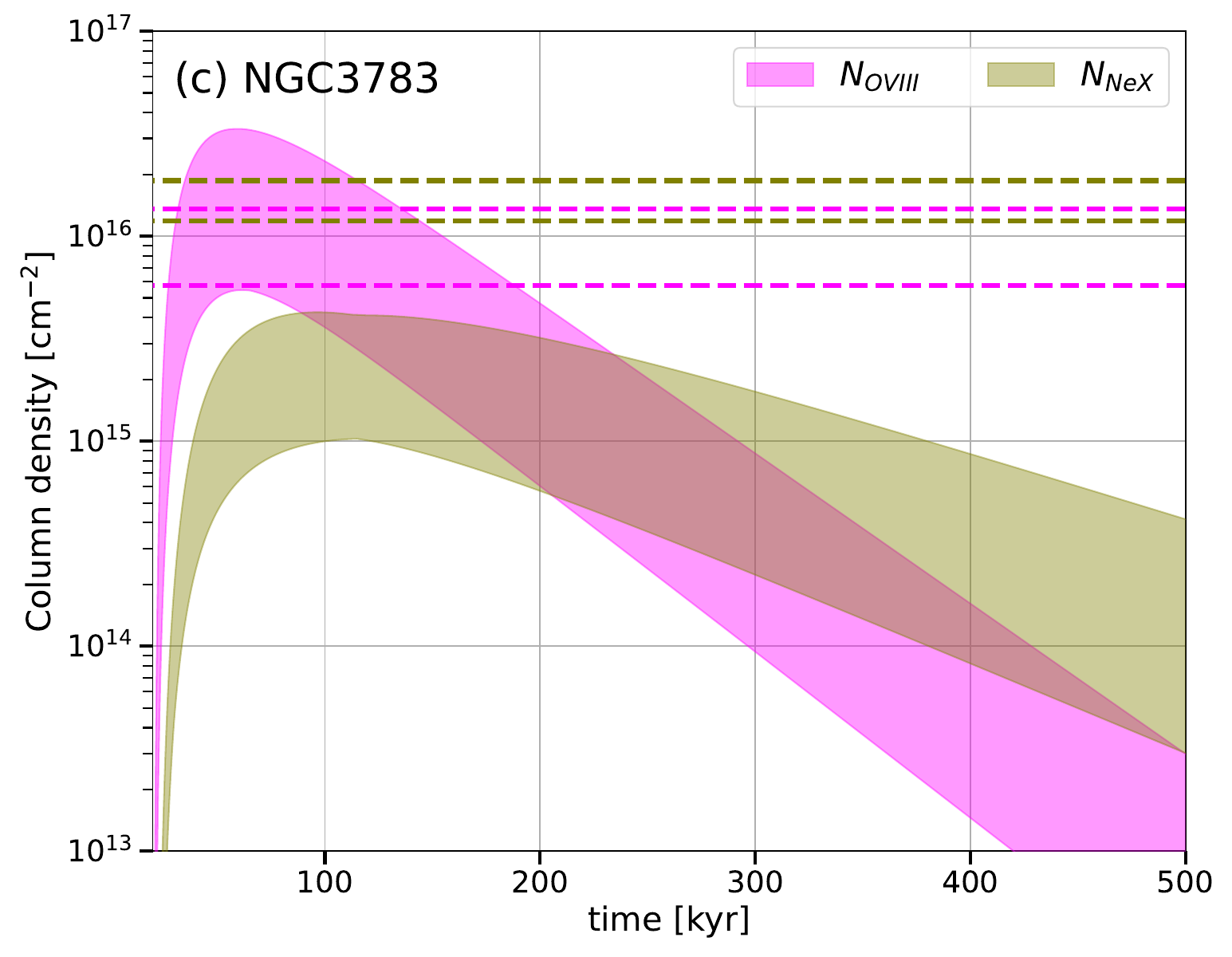}
    
    \caption{Column density of ions as a function of time for three lines of sight. The {\it top (a), middle (b)} and {\it bottom (c)} panel shows the column density towards Mrk421 (\citet{Das2021}), IES1553+13 (\citet{Das2019a}), and NGC3783 (\citet{McClain2023}) respectively. The magenta, green, and blue bands show the column density range of OVIII, NeX, and SiXIV respectively calculated from our model. The dashed lines in the same color show the upper and lower limit of the observed column density of respective ions.}
    \label{Fig:ion_column}
\end{figure}

Since we are interested in H-like, i.e., almost fully ionized ions, we fix the value of $n_e$ of the gas assuming it is almost fully ionized. This does not affect the evolution of highly ionized atoms but allows us to readily calculate the fractional abundance of any ion for different atoms independent of the composition.  We adopt $n_e=0.01$ cm$^{-3}$ for Oxygen, and Neon and $n_e=0.005$ cm$^{-3}$ for Silicon, corresponding to almost fully ionized gas for the estimated values of $n_{\rm ion}$ from simulations. The resulting values of ion fractions of an element are computed for the range of temperatures mentioned above yielding a range of ion fractions for each ion.  
This range of ion fractions for each ion of an element when folded with the range of total elemental column densities found in various density profiles, results in a band of ion column density as a function of time. 
We consider total column densities of $(1-6)\times 10^{16}$, $(2-8)\times 10^{15}$, and $(1-5)\times 10^{15}\,{\rm cm^{-2}}$ for Oxygen, Neon, and Silicon, respectively, corresponding to the range of values found in  all the density profiles. Figure \ref{Fig:ion_column} shows the calculated band of column density of OVIII, NeX and SiXIV, superposed with the observed column densities of respective ions towards Mrk421, IES1553+113 and NGC3783 (see Table \ref{tab:data}). The dashed lines in the respective colors show the observed upper and lower limits. The curved bands show the ionic column densities from our model.

\subsubsection{Mrk 421}
The top panel of Figure \ref{Fig:ion_column} compares the results for our fiducial case with the observations towards Mrk 421 \citet{Das2021}. Interestingly, the colored bands of our model prediction for all three ions (OVIII, NeX and SiXIV) show an overlap at $\sim 200\hbox{--}300$ kyr, which also approximately match the observed bands for OVIII and NeX. There is a slight shortfall in the SiXIV column density, by a factor of $\approx 2$, which can be potentially explained by variations in supernova yields.

\subsubsection{IES1553+113}
The middle panel of Figure \ref{Fig:ion_column} shows the comparison with observed column densities of OVIII and NeX towards IES1553+113 \citet{Das2019a}. We again find an overlapping region for OVIII and NeX column densities, around $t\sim 100\hbox{--}400$ kyr, and it matches the observed column density of OVIII to a large extent. The predicted NeX column density falls short of the observed values, again by a factor $\approx 2$, which can be due to uncertainties in SN yield.

\subsubsection{NGC 3783}
We show the comparison of our results with the observations towards NGC 3783 \citet{McClain2023} in the bottom panel of Figure \ref{Fig:ion_column}. Again, there is an overlap between the column densities of OVIII and NeX for similar timescales as in the previous case. In this case, the OVIII column densities can explain the observed values, and for NeX as well, if the yield is increased by a factor of $\approx 2$ or the ambient density is increased to $\approx 0.3$ cm$^{-3}$.

\section{Discussion}\label{sec:discussion}
We have shown that the observed column density of OVIII, NeX and SiXIV 
can be naturally explained if the line of sight passes through the reverse shocked gas in a non-radiative SNR in the extra-planar region of MW within $\sim 100$--$400\,{\rm kyr}$ from the time of the explosion. We have explored the variation of column density and temperature with reasonable values of ejecta mass, blast energy, and ambient density. As mentioned in the introduction, such supernova explosions far from the Galactic plane can naturally be explained by massive runaway stars. However, to evaluate the feasibility of runaway massive stars exploding as supernovae above the Galactic plane as the primary source for the observed events, it is important to compare the frequency of such events with the observed frequency. In the solar neighbourhood, among massive stars that end their life as CCSNe, i.e., O stars and some B stars, $\sim 30\%$ of O stars and $\sim 10\%$ of B stars are found to be runaway stars \citep{gies1986,stone1991}. On the other hand, the frequency of Galactic CCSN within the last $\lesssim 1$ Myr is constrained to be $ 1.9\pm1.1$ per century from the observed $\gamma$-ray emission from $^{26}$Al \citep{Diehl2006}. Based on the results presented in Section~\ref{subsec:comp_obs}, the observed column densities of OVIII, NeX, and SiXIV can be explained if the explosion occurred $\sim 100$--$400\,{\rm kyr}$ ago. Assuming $\sim 20\%$ of all CCSNe are due to runaway  OB stars with a total rate of $\sim 2$ per century, the number of such progenitors that were born during an interval of $\Delta \sim 250\,{\rm kyr}$ is $N_{\rm runaway}(\Delta)\sim 1000$. 

In order to estimate the final location of the supernova from runaway stars, the motion of the star after it is ejected from the Galactic disk needs to be simulated. We use \textsc{galpy} \citep{galpy} to simulate the motion of runaway stars within the Galactic potential. Runaway OB stars can have large ejection velocities of up to $\sim 400\,{\rm km \, s^{-1}}$ from the disc that can be modelled by a Maxwellian distribution as found in \citet{Silva2011}. Following their result, we adopt a Maxwellian velocity distribution with a mean velocity of  $150\,{\rm km \, s^{-1}}$ for the ejection velocity from the Galactic disk whose direction is assumed to be isotropic. Each runaway star is considered to have travelled a period of $\tau$ before it explodes as a supernova. The runaway star can be ejected either during the early phase of its life via the dynamical ejection mechanism from young open clusters \citep{poveda1967} or during the later phase via the binary ejection mechanism when one of the stars in a binary explodes as a supernova \citep{blaauw1961}. We choose the value of $\tau$ from a uniform distribution ranging from $1$--$10\, {\rm Myr}$ to account for both early and late ejection. The initial radial coordinate of birth location is assumed to follow the stellar mass distribution from \citet{mcmillan2017}. We use \texttt{MWPotential2014} in \textsc{galpy} as the model for the MW potential with the circular velocity $v_{\rm c}(R_\odot = 8.21 \,{\rm kpc}) = 233.1\,{\rm km \, s^{-1}}$ adapted from the best-fit values for the MW from \citet{mcmillan2017}. 

\begin{figure}
    \centering
    \includegraphics[width=\columnwidth]{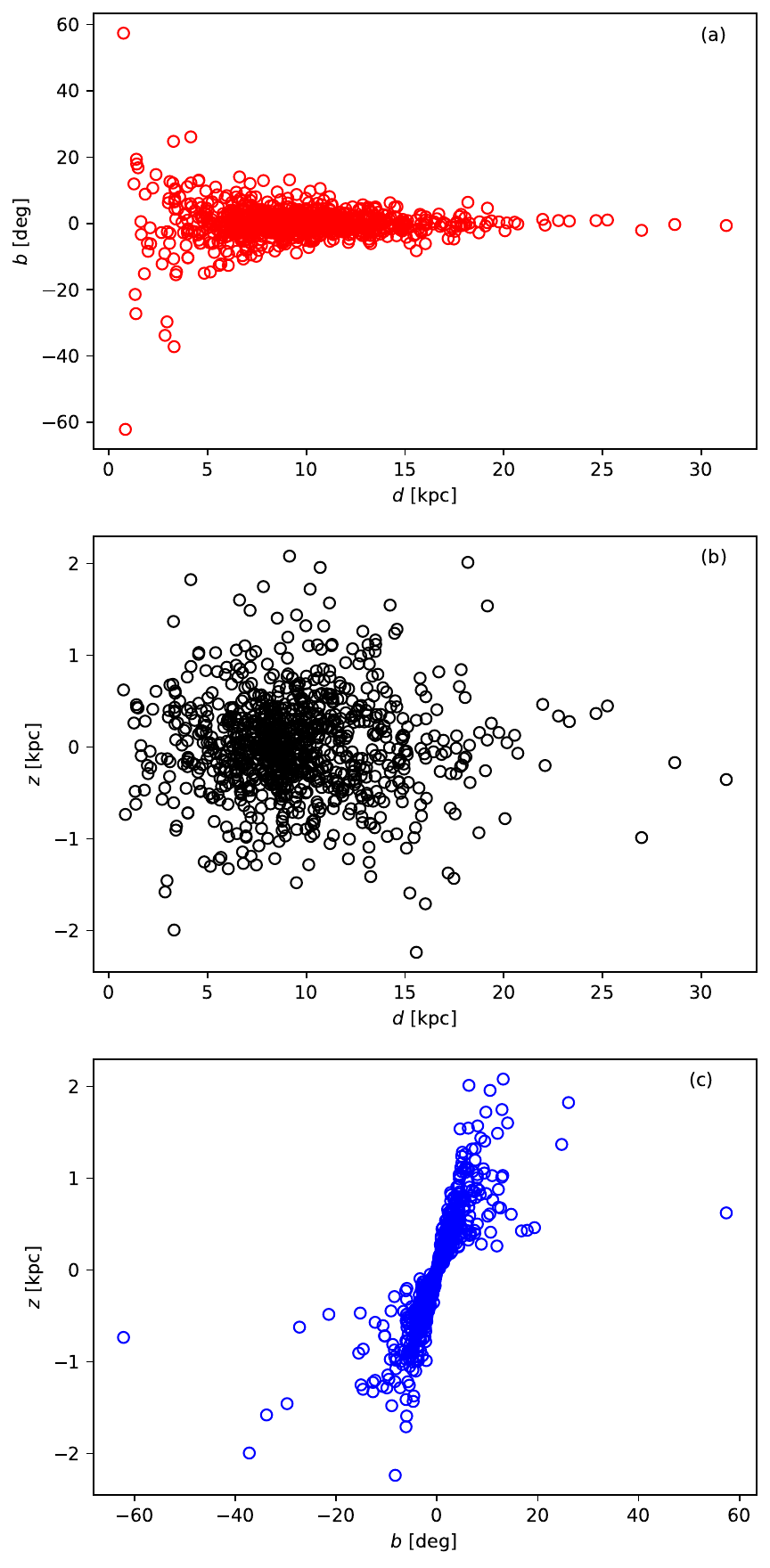}
    \caption{The final position of 1000 runaway massive stars where they explode as core-collapse supernovae: (a) latitude $b$ versus distance $d$ from the Sun, (b) Height $z$ from the Galactic plane versus $d$, and (c) $z$ versus $b$. }
    \label{fig:galpy1}
\end{figure}

Figure~\ref{fig:galpy1}a-b show the final location of supernova explosion for a given realization of 1000 runaway stars in terms of their latitude $b$  and height $z$ above the Galactic plane as a function of their distance $d$ from the Sun. Fig.~\ref{fig:galpy1}c shows the corresponding locations in the $z$-$b$ plane. As can be seen from Fig.~\ref{fig:galpy1}b, a majority ($\sim 93\%$) of the supernovae from runaway stars are confined to $|z|\leq 1\,{\rm kpc}$ and almost all of them at $|z|\lesssim 2\,{\rm kpc}$ with $d \lesssim 25\,{\rm kpc}$. Because of the low value of $z$, the overwhelming majority of $\sim 98.5\%$ of the sources have $|b|\lesssim 15^{\circ}$ where the sources with lower values of $d$ tend to have higher $b$. 
However, there are still $\sim 15$ events that occur at high latitudes of $> 15^{\circ}$ with values of up to $\sim 60^{\circ}$, out of which $\sim 9$ and $\sim 5$ occur at low heights of $|z|\lesssim 1\,{\rm kpc}$ and $|z|\lesssim 0.5\,{\rm kpc}$, respectively. These low $z$ and high $b$ sources are ideally suited to be the ones that correspond to observed sources that can satisfy the observed high values of $b$ along with high ambient densities. The cold neutral medium (CNM) and warm neutral medium (WNM) have mid-plane densities of $\approx 0.3$ and $0.1\,{\rm cm^{-3}}$, with a scale height of $\approx 0.15$ and $0.40\,{\rm kpc}$, respectively \citep{kalberla2009}. Thus, WNM in particular can provide the assumed fiducial ambient density of $0.1\,{\rm cm^{-3}}$. As a consequence of low $|z|\lesssim 1\,{\rm kpc}$, sources with high latitudes are also constrained to be confined to relatively low values of $d$ ranging from $\sim 0.75$--$1.5\,{\rm kpc}$. In particular, sources at highest latitudes of $\sim 60^{\circ}$ are also the closest sources at $d\sim 0.75$--$0.83\,{\rm kpc}$. 

We note that recently, \citet{Lara-DI2023} have reported very high average column density of SiXIV of $(8.7\pm1.6)\times 10^{15}\,{\rm cm^{-2}}$, implying supervirial gas with $T\gtrsim 10^7\, {\rm K}$, using stacked spectra from \textit{Chandra} along 46 different Quasar sightlines. We note here that the `average' column density of SiXIV reported by \citet{Lara-DI2023} is that obtained from the stacked `line', and is not the total column density divided by the number of sightlines used in stacking.
A follow-up work found three distinct phases similar to \citet{Das2019a}, including a very hot phase with $T\sim 3.2\times 10^{7}\,{\rm K}$ from the same stacked sources \citep{Lara-DI2024}.  Even though these results do provide hints that supervirial gas with high column densities for H-like ions like SiXIV is possibly widespread in the CGM, currently, there is no robust observational constraint on the covering fraction of the absorbing hot gas. Future detections of high column density along a large number of individual sightlines are required to determine the covering fraction. In this regard, it is important to note here that $10^7$ K gas seen in emission needs to be differentiated from that seen in absorption, and although the covering fraction of emitting `super-virial' gas is large, the case of the gas seen in absorption is likely different.

With regard to the scenario presented here, assuming the size of shocked gas from each SNR to be $\sim 30\,{\rm pc}$, the covering fraction for high latitude ($b>15^\circ$) sources ranges from $\sim 0.015\hbox{--}0.030$ steradian corresponding to $\lesssim 0.25\%$ of the sky whereas the covering fraction for low latitude ($b<15^\circ$) sources ranges from $0.05\hbox{--}0.07$ steradian corresponding $\lesssim 0.7\%$ of the sky. Thus, although the current scenario of CCSN from runaway stars can explain the high column density of highly ionized H-like ions like SiXIV along individual sightlines, they cannot account for a widespread CGM-wide distribution. It is important to note that regardless of whether superviral gas is present along particular sightlines or spread throughout the Galaxy, SNR from massive runaway stars will be present along multiple sightlines including high altitudes that would have high column density of highly ionized H-like ions.  
In this regard, it is interesting to note that at least  7 and 17 sources out of the total 46 sources in \citet{Lara-DI2023} have $|b|<20^{\circ}$ and $|b|<30^{\circ}$, respectively. Some of these sources could be due to CCSN from runaway massive stars.

Besides runaway stars, there are also other types of stars above the Galactic disk, although their numbers are relatively much smaller. Binary stars through dynamical interaction with super-massive black hole (SMBH) in the center of MW can also eject one of the stars with a speed more than the escape velocity of MW, thus producing hyper-velocity stars (\citet{Brown2015}). Some of the hypervelocity stars in the extra-planar region might have an external origin as well (\citet{Gulzow2023}). Apart from being launched into the extra-planar region, stars are also formed {\it in situ} above the disk (\citet{Tuellmann2003}, \citet{Martinez2021}, \citet{Howk2018}). We do not consider these categories of stars important for our discussion because of their small numbers.

We should also like to mention the possibility of the ambient density being comparable to or even larger than the fiducial value of $0.1$ cm$^{-3}$ at a height $\ge 0.5$ kpc above the disk.
It is known that there are HI clouds above the disk, up to a height of $\sim 1.5$ kpc, and with average densities $\approx 0.3$ cm$^{-3}$ and diameters of $\approx 20\hbox{--}30$ pc \cite{Lockman2002}. The current understanding is that these clouds originate in the disk and have moved up owing to their large vertical velocities. They also appear to have large turbulent speed, of order $\approx 75$ km s$^{-1}$ \cite{Kalberla2008}, which is estimated from fitting the emission spectrum. According to \cite{Lockman2002}, as much as half the mass of the HI halo of the MW could be contained in such `halo clumps'. 
These clouds are likely to be supported against the gravitational potential by random turbulent speeds, with which they can reach a height up to a few kpc \cite{kalberla2009}. It is difficult to estimate the volume filling factor of such clouds, from their observed incidence rate (on average, one cloud per two square degrees) because of projection effects. 
However, the observed gas densities and the typical heights of these clouds are within the ranges required in our scenario. It is possible that the SNe discussed here could go off inside one such clump above the disk, in which case our assumption of ambient density would be quite reasonable.

Can these SNe be detected by other means, or do they leave any other signature? SNe produce radioactive $^{26}$Al which  $\beta$-decays to excited states of $^{26}$Mg that subsequently emits the well known 1.809 MeV $\gamma$-ray \citep{Diehl2006}. The yield of $^{26}$Al in massive stars of $10$--$30\,\Msun$ range from $\sim (0.3$--$1.5)\times 10^{-4}\,\Msun$ \citep{tur2010}. This corresponds to a photon flux of $F_\gamma\lesssim 1.8\times10^{-6}\,\gamma\,{\rm cm^{-2}\,s^{-1}}$ for a source at a distance of  $1\,{\rm kpc}$. This is more than an order of magnitude less than the line sensitivity of INTEGRAL at $1\,{\rm MeV}$ of $2.4\times10^{-5}\,\gamma\,{\rm cm^{-2}\,s^{-1}}$. This implies that even for the highest latitude sources that are also the nearest at $d\sim 1\,{\rm kpc}$, INTEGRAL cannot detect $\gamma$-rays from $^{26}$Al. However, future proposed $\gamma$-ray detectors such as GRAMS \citep{aramaki2020}, that has a proposed line sensitivity of $(1.3\hbox{--}7.2)\times 10^{-7}\,\gamma\,{\rm cm^{-2}\,s^{-1}}$ for the $1.809\,{\rm MeV}$ $\gamma$-ray, would have a very good chance of detecting such sources.

The radio luminosity of SNRs at, say, $1.4$ GHz can be estimated, using Equation 2 of \cite{leahy2022}, as,
\begin{eqnarray}
\nonumber
    L_{1.4}  && \approx 1.7 \times 10^{22} \, {\rm erg}\, {\rm s}^{-1}\, {\rm Hz}^{-1} \\
    && \times \, \Bigl ({\eta \over 5 \times 10^{-3}} \Bigr ) \, \Bigl ({R_b \over 20 \, {\rm pc}} \Bigr )^3 \, \Bigl ( { v_b \over 300 \, {\rm km/s}} \Bigr )^{3.2} \,
\end{eqnarray}
where $R_b, v_b$ are forward shock distance and velocity, and $\eta$ is a parameter that encodes various physical processes, such as the fraction of supernovae energy going into relativistic electrons, the efficiency of magnetic field amplification, etc. For our fiducial case, the forward shock has a speed of $\approx 300$ km s$^{-1}$ when the forward shock reaches $\sim 20$ pc. \cite{leahy2022} have found, after comparison with observations, the value of $\eta$ to lie between $10^{-5}\hbox{--}10^{-1}$. Its value is likely to be on the lower side, given the weak magnetic field expected above the Galactic disc. If we require the radio flux at $1.4$ GHz from an SNR at a distance of $1$ kpc to be comparable to the observed flux of, say, Mrk 421 \citet{mrk421-2005} ($\nu F_\nu \approx 5 \times 10^{-14}$ erg cm$^{-2}$), then the value of $\eta\approx 2 \times 10^{-5}$, which is the acceptable range, as mentioned above. 
In other words, the radio signature of the proposed SNR would be indistinguishable from or swamped by the background source, provided the supernova goes off at a distance of a kpc. However, at lower frequencies, while the synchrotron emission from SNR is expected to have a spectral index of $\sim -0.5$, that of the background blazar typically has a flatter index (with an average radio spectral index of $0.03$, see Fig.~25 of \citet{abdo2010})
, that extends down to $\sim 300$ MHz \citet{Nori2014}, below which the index becomes positive due to self-absorption \citep{Mooney2019}. Therefore, a low-frequency observation of these blazars may be able to pick up the emission from the foreground SNR, being brighter than the blazar at these frequencies. However, the radio spectral indices of blazars vary, and so do those of SNRs, from $\sim -0.85$ (shell-type) to $\sim 0$ (filled-center type) (\citet{Ranasinghe2023}, their Fig. 7), and such variations may make it difficult to detect a foreground SNR in radio.

Finally, we note that a key difference between the scenario proposed in this work from the analysis used to infer super-virial gas in studies such as \citet{Das2019b,Das2021, Lara-DI2024}, is the assumption of CIE. As mentioned in Section~\ref{sec:timesclaes}, the usual assumption of CIE is not at all applicable for the ejecta heated by reverse shock described in this work. The abundance of ions is not in equilibrium but changes with time due to ionization and eventually all atoms become fully ionized. However, the timescale for ionization is different for each element with longer timescales for heavier atoms. In our scenario, although a temperature cannot be assigned from an observed absorption line due to lack of CIE, the super-viral nature of the gas is directly implied from large column densities observed for highly ionized H-like ions such as OVIII, NeX, and SiXIV.

\section{Summary}\label{sec:summary}
Observations of the CGM of the MW in absorption along three lines of sight indicate rather large values of the column density of highly ionized H-like ions (Table \ref{tab:data}) with temperatures larger than the virial temperature of MW.
In this paper, we argue that the observed column density of H-like ions such as OVIII, NeX, and SiXIV implying super-virial temperatures exceeding a few million degrees K can be naturally produced by supernovae remnants red{from runaway stars} above the Galactic disk, because of heating of metal-enriched ejecta by reverse shock. 
We show that observed column density is achievable with reasonable values of ejecta mass of $\rm 10 \, M_\odot$, blast energy of $10^{51}$ erg, and the ambient density of $\sim 0.1$ cm$^{-3}$.
We would like to propose that if the `super-virial' phase with high column density of H-like ions is found to be limited to a few lines of sight rather than being widespread throughout the CGM, it is likely caused by the chance coincidence of supernovae remnants of runaway stars in the lines of sight to background sources.


\section*{Data Availability}

Data is available upon reasonable request.

\section*{Acknowledgements}

We thank anonymous referee for the constructive comments. We thank Smita Mathur for the useful discussion on observational aspects of the `super-virial' gas. YS acknowledges the hospitality of RRI.

%






\bibliography{reference}{}
\bibliographystyle{aasjournal}



\end{document}